\documentstyle[preprint,aps]{revtex}
\begin{document}

\newcommand{\be}{\begin{equation}}
\newcommand{\ee}{\end{equation}}
\newcommand{\beann}{\begin{eqnarray*}}
\newcommand{\eeann}{\end{eqnarray*}}
\newcommand{\bea}{\begin{eqnarray}}
\newcommand{\eea}{\end{eqnarray}}
\newcommand{\nn}{\nonumber}
\newtheorem{df}{Definition}
\newtheorem{thm}{Theorem}
\newtheorem{lem}{Lemma}

\draft

\title{Relation between the guessed and the derived
  super-Hamiltonians for spherically symmetric shells}

\author{Petr H\'{a}j\'{\i}\v{c}ek} 
\address{Institute for Theoretical Physics,
University of Bern,
Sidlerstrasse 5, CH-3012 Bern, Switzerland}

\date{April 1998}

\maketitle

\begin{abstract}
  The Hamiltonian dynamics of spherically symmetric massive thin shells in the
  general relativity is studied. Two different constraint dynamical systems
  representing this dynamics have been described recently; the relation of
  these two systems is investigated. The symmetry groups of both systems are
  found. New variables are used, which among other things simplify the
  complicated system a great deal. The systems are reduced to presymplectic
  manifolds $\Gamma_1$ and $\Gamma_2$, lest non-physical aspects like gauge
  fixings or embeddings in extended phase spaces complicate the line of
  reasoning. The following facts are shown.  $\Gamma_1$ is three- and
  $\Gamma_2$ is five-dimensional; the description of the shell dynamics by
  $\Gamma_1$ is incomplete so that some measurable properties of the shell
  cannot be predicted. $\Gamma_1$ is locally equivalent to a subsystem of
  $\Gamma_2$ and the corresponding local morphisms are not unique, due to the
  large symmetry group of $\Gamma_2$. Some consequences for the recent
  extensions of the quantum shell dynamics through the singularity are
  discussed. 

\end{abstract}


\section{Introduction}
The dominating physical problem of the theory of gravitation is the
gravitational collapse and the inevitable singularity. That is the point,
where the classical theory breaks down; one expects that the quantum theory
will help.

The project of which the present paper is a part focuses on simplified models
of gravitating systems, whose quantum mechnics can be constructed without much
technical and conceptual difficulty. We hope that such models can help us to
find quantization methods that are, in short, a) gauge (reparametrization)
invariant and b) liberated from semiclassical thinking. Indeed, the gauge
invariance is an issue, because a gauge fixation leads much more easily to
gauge-dependent results in quantum gravity than, say, in the Yang-Mills field
theory. This has to do with the so-called {\em problem of time} \cite{KKcan},
\cite{CI}. The term `semi-classical thinking' is to express uneasiness about
the extensions to the gravidynamics of the naive assumption that every
dynamics is a theory of motion of some objects in some spacetime (cf.\ 
`covariant quantization' \cite{DW}, string theory \cite{GSW}, or `effective
field theory' \cite{D}). As the spacetime is itself dynamical in the presence
of gravitation, the only method complying with the assumption seems to be an
expansion around a classical solution, and so it must be, it seems at least, a
sort of WKB approximation.

Our pet model is the spherically symmetric thin shell of dust in the general
relativity. In refs.\ \cite{H1}, \cite{HKK}, \cite{H2} and \cite{B-H}, a
sufficiently simple super-Hamiltonian for this system was guessed from the
equations of motion, so that the model could be readily quantized. We shall
call this method the {\em Warsaw} approach\footnote{A systematic and general
  exposition of this method (Ref.\ \cite{B-H}) was first given at a Banach
  Center Workshop, Warsaw.}. 
This quantum theory lead to some unexpected results:
\begin{enumerate}
\item Existence of a unitary scattering theory; after a contracting phase, the
  shell goes through a more or less probable intermediate state of black and
  white hole and then spacetime wave packet expands again; the probability of
  the hole stage depends on the energy of the shell.
\item This unitary regime exists for all values of energy but only if the
  total rest mass does not exceed the Planck mass; what happens with more
  massive shells remains unclear.
\item There is no spacetime geometry that can be associated with any
  particular scattering process: the wave packets contain linear combinations
  of wave functions that describe spacetimes with contracting shells and black
  holes as well as spacetimes with expanding shells and a white holes.
\end{enumerate}

The method, however, has several weak points.
\begin{enumerate}
\item The way from a dynamical equation to a Hamiltonian principle is
  non-unique and it is unclear how much the final quantum results depend on a
  particular guess. Some kind of uniqueness has been shown in Ref.\
  \cite{B-H}, but this is not yet completely reassuring, as we shall also
  see in the present paper.
\item The great simplicity of the super-Hamiltonian is achieved by a very
  special choice of variables: they are the coordinates of the shell in the
  subspacetime at one side of the shell. This subspacetime of description
  (left subspacetime) lies on the opposite side of the shell than the
  subspacetime where the observers spend most of their time (right
  subspacetime). The drawback is 
  that some observable properties of the shell are not contained among (or
  calculable from) these variables. A controversy arises as to whether and
  where the observers will see the expanding shell.  Undetermined are for
  example the scattering time delays of the expanding packet with respect to
  the contracting one (usually given by the derivatives with respect to energy
  of the phase shifts of the S-matrix), which should in principle be
  measurable. The possibility was even discussed that the wave packet expanded
  into a different right subspacetime than where it originally contracted so 
  that the scattering time delays made no sense.
\item The method of self-adjoint extension of Hamiltonian operator was applied
  in Refs.\ \cite{H1}, \cite{HKK}, and \cite{H2} to make the quantum evolution
  complete (unitary). This method seemed to some physicists too formal and
  suspicious to cope satisfactorily with the problem of the singularity.
\end{enumerate}

The present paper is an attempt to deal with all three problems. The analysis
will be based on Refs.\ \cite{H-K} and \cite{H}; we shall refer to this
approach as the Potsdam one\footnote{Ref.\ \cite{H} was written down at the
  Albert Einstein Institute, Potsdam.}. The Potsdam Hamiltonian action for
spherically symmetric massive shells has been derived from first principles,
so one can say something about the first problem. The Postdam dynamics of the
shell is described by its coordinates in both subspacetimes left and right
from the shell, and so it contains complete information about the motion. This
enables us to say something about the second problem. Finally, we shall find
that the method of self-adjoint extension is incomplete in certain respect,
and so there is some information about the third problem, too. There is,
of course, a strong motivation for the Warsaw approach: this is the etremely
simple form of its action.

The plan of the paper is as follows. In Sec.\ \ref{sec:descript}, we shall
briefly collect the results of Refs.\ \cite{B-H} and \cite{H} that will be
needed, so that the paper becomes self-contained. New is the systematic use of
the Kruskal coordinates and more or less complete description of the
symmetries that the systems admit. Sec.\ \ref{sec:cartan} is based on the idea
that the `naked' or minimal mathematical structure that underlies a given
constrained system is the constraint manifold together with the induced
presymplectic form. According to this idea, two constrained systems are
equivalent, if they define the same presymplectic manifolds. This notion is
completely gauge (and reparametrization) invariant; it is moreover independent
of the details of imbedding of the constraint manifold in an extended phase
space, which is not physical and not unique. In Sec.\ \ref{sec:cartan}, we
reduce the Warsaw and the Potsdam descriptions to these `naked' forms. The
Warsaw presymplectic manifold $(\Gamma_1,\Theta_1)$ is three-dimensional and
the Potsdam presymplectic manifold $(\Gamma_2,\Theta_2)$ is five-dimensional.
As a byproduct of the transformations made in this section, `miraculous' new
variables are found which considerably simplify the Potsdam action.

In Sec.\ \ref{sec:excl}, we pick up a three-dimensional subsystem
$(\Gamma_E,\Theta_E)$ from the Potsdam system $(\Gamma_2,\Theta_2)$ that has a
chance to be equivalent to a particular Warsaw system. Concretely, the
Schwarzschild mass of the left subspacetime must be fixed, and the
corresponding cyclic coordinate must be excluded. Sec.\ \ref{sec:trans} is
devoted to the search of a morphism of presymplectic manifolds that would map
$(\Gamma_1,\Theta_1)$ onto $(\Gamma_E,\Theta_E)$. We set up a partial
differential equation for this map and solve it in suitable coordinates. A
careful study of the results reveals that
\begin{enumerate}
\item the two presymplectic manifolds are {\em locally} but not {\em globally}
  equivalent, and
\item the local equivalence maps (morphisms of presymplectic manifolds) are
  not unique. 
\end{enumerate}
In more detail, the manifolds $\Gamma_1$ and $\Gamma_E$ can be covered with
open patches such that each one on $\Gamma_1$ is equivalent to one on
$\Gamma_E$, but the map cannot be extended to the outside of the patch,
because it diverges at the boundary. Scattering trajectories contracting from
right (left) and expanding to right (left) can both never lie within one and
the same patch. Where two patches overlap, the two corresponding maps differ
by a symmetry. The huge amount of symmetry in the Potsdam approach is the
cause for the non-uniqueness of the map.

Let us dwell a little more on these results. The local equivalence explains
why the same radial equation results from both systems.  What is, however, a
possible physical significance of just local but not global equivalence of
constraint dynamical systems remains unclear. On could imagine, for example,
pasting together several copies of $\Gamma_1$ and $\Gamma_E$ using the local
equivalence maps. The result, however, seems to be a (possibly non-Hausdorff)
presymplectic manifold with no reasonable physical interpretation (this is
discussed at the end of Sec.\ \ref{sec:match}).

It is also conceivable that the global inequivalence of the systems is not
important for the quantum theory; the delicate points where the map can
diverge, lie all at the boundary between the scattering and the bound
trajectories. One can speculate that this discontinuity is not relevant for
the quantum mechanics, because the bound states become discrete anyway. If we
cut out this less dimensional boundary from both classical systems, we obtain
systems that {\em are} globally equivalent. Further study is necessary.

Suppose next that some form of weakened equivalence between the two systems
makes sense. Then, given an equivalence map, we can consider the Warsaw system
as a part of the Potsdam one, and the missing information about the position
of the shell in the right subspacetime can be provided. The Warsaw variables
can be regarded as coordinates on a part of the Potsdam system. In fact,
it turns out then that the Warsaw variables are coordinates of the shell in
the left subspacetime only according to their name; in reality, they play the
role of coordinates in the right subspacetime.  Hence, the position of the
shell in the right subspacetime is well-determined; paradoxically, it is the
position of the shell in the left subspacetime that is uncertain (see the
discussion at the beginning of Sec.\ \ref{sec:trans}).

Applying this point of view to the quantum mechanics of Refs.\ \cite{H1} and
\cite{HKK}, we can make some progress. The self-adjoint extension of the
Hamiltonian defines a unitary dynamics in the Warsaw coordinates of a part of
the Potsdam system.  It follows that each wave packet reemerges, during the
expanding part of this unitary evolution, in the same right subspacetime in
which it originally started to contract, because the variables describing it
all the time are coordinates on this subspacetime. This is discussed in more
detail at the beginning of Sec.\ \ref{sec:trans}.

On the other hand, the non-uniqueness of the equivalence map has an
unpleasant consequence: although the self-adjoint extension of the dynamics
seems to be unique, it is so only with respect to a particular choice of
Warsaw variables. Different equivalence maps lead to different choices of the
variables and these, in turn, to different dynamics of the Potsdam
system. This is explained at the beginning of Sec.\ \ref{sec:trans}.
The difference is measurable, because the resulting scattering time delays are
different. This suggests that one either has to look for some additional
principle that could, together with the self-adjoint extension, lead to a
unique set of time delays, or to look for some interpretation of the
non-uniqueness (like, say, a loss of information) or to look for another way
of dealing with the singularity. Future research may clarify the point.

\section{Description of the shell dynamics}
\label{sec:descript}
In this section, we briefly collect and round off some results scattered in
literature making so the paper self-contained.

\subsection{The shell spacetime}
A sherically symmetric thin-shell spacetime solution of Einstein equations
will be described in this subsection following closely Ref.\ \cite{B-H}.
Consider two Kruskal spacetimes ${\mathcal M}_1$ and ${\mathcal M}_2$ with
Schwarzschild masses $E_1$ and $E_2$. Let $\Sigma_1$ be a timelike
hypersurface in ${\mathcal M}_1$ and $\Sigma_2$ be one in ${\mathcal M}_2$.
Let $\Sigma_1$ divides ${\mathcal M}_1$ into two subspacetimes, ${\mathcal
  M}_{1+}$ and ${\mathcal M}_{1-}$, and similarly $\Sigma_2$ divides
${\mathcal M}_2$ into ${\mathcal M}_{2+}$ and ${\mathcal M}_{2-}$; we chose
fixed time and space orientation in the two-dimensional Kruskal spacetimes so
that future and past as well as right and left are unambiguous; let then
${\mathcal M}_{2+}$ and ${\mathcal M}_{1+}$ be right with respect to
${\mathcal M}_{2-}$ and ${\mathcal M}_{1-}$. Let $\Sigma_1$ and $\Sigma_2$ be
isometric; then the spacetime ${\mathcal M}_{1-}$ can be pasted together with
the spacetime ${\mathcal M}_{2+}$ along the boundaries $\Sigma_1$ and
$\Sigma_2$. The result is a shell spacetime ${\mathcal M}_s$. As everything is
spherically symmetric, only the two-dimensional Kruskal spacetimes are
relevant.

The observers are assumed to be in the asymptotic region of
${\mathcal M}_{2+}$. Given a shell spacetime, we shall often leave out the
indices 1 and 2, having right (left) energy $E_+$ ($E_-$), shell
trajectory $\Sigma$, and the right (left) subspacetime ${\mathcal
  M}_{+}$ (${\mathcal M}_{-}$). Thus, ${\mathcal M}_{+} = {\mathcal M}_{2}
\cap {\mathcal M}_{s}$ and ${\mathcal M}_{-} = {\mathcal M}_{1}
\cap {\mathcal M}_{s}$. 

The three-dimensional shell surface $\Sigma$ carries the energy-momentum
tensor $T^{kl}$ of the shell. This will be assumed in the form of ideal fluid
\[
  T^{kl} = (\rho + p)T^kT^l + p\gamma^{kl},
\]
where $\rho$ is the surface mass density, $p$ the negative surface tension,
$T^k$ a unite timelike vector field (the three-velocity of the fluid)
tangential to $\Sigma$, and $\gamma_{kl}$ is the metric induced on the shell
from the surrounding spacetime. Let $p = p(\rho)$ be the equation of state.

The spherical symmetry and the energy-momentum conservation lead to the
matter equation
\[
  {\mathcal A}\frac{d\rho}{d{\mathcal A}} + \rho +p(\rho) = 0,
\]
where ${\mathcal A}$ is the surface of the shell, ${\mathcal A} := 4\pi
R^2$. We choose one of the particular solutions of the matter equation,
$\rho({\mathcal A})$, and define the so-called mass function $M(R)$ by
\[
  M(R) := {\mathcal A}(R)\rho({\mathcal A}(R)).
\]
For example, the dust equation of state, $p = 0$, implies that $M(R) =$ const,
and each value of the constant defines a particular solution.

The spacetime around the shell already satisfies the Einstein equations; thus,
the only non-trivial equation still to be satisfied is the jump condition, the
so-called Israel equation \cite{I}. It implies (for a derivation, see
Ref.\ \cite{B-H}) the following two 
equations for the embedding functions $T_{\epsilon}(s)$ and $R(s)$ of the
shell in ${\mathcal M}_\epsilon$, where $\epsilon$ is a sign,
$\epsilon = \pm 1$ and $s$ is the proper time along the radial generators
of the surface. \\
A) The radial equation:
\be
 \dot{R}^2 + V(R) = 0,
\label{radeq}
\ee
where
\be
  V(R) := -\frac{M^2(R)}{4R^2} - \frac{E_+ + E_-}{R} - \frac{(E_+ -
  E_-)^2}{M^2(R)} + 1,
\label{potent}
\ee
and \\
B) the time equation (valid for all future oriented shell motions)
\be
  \dot{T}_{\epsilon} = \frac{1}{M(R)F_\epsilon}\left(E_+ - E_- -
    \epsilon\frac{M^2(R)}{2R}\right), 
\label{timeq}
\ee
where
\[
  F_\epsilon := 1 - \frac{2E_\epsilon}{R}
\]
is the Schwarzschild function.

Here, $T_\pm$ and $R$ are the Schwarzschild coordinates of the shell in
${\mathcal M}_\pm$ (they are of course singular at the four horizons of the two
spacetimes ${\mathcal M}_\pm$)

There are two types of solutions: bound and scattering; the scattering
trajectories can be divided into {\em expanding} ($\dot{R}>0$) to the right
and to the left, and {\em contracting} ($\dot{R}<0$) from the right
and from the left (the usual notions of out- and in-going are not adequate to
dynamics in the Kruskal spacetime). We stress that the four possibilities are
unambiguous (see Ref.\ \cite{B-H}): for each (scattering) shell spacetime,
only one of these is realized in {\em both} subspacetimes ${\mathcal M}_\pm$
simultaneously.

\subsection{Warsaw approach}
In this subsection, we describe a short-cut approach (cf.\ e.g.\ 
\cite{HKK} and \cite{B-H}) to the shell dynamics. The first step of the
approach is that only a subclass of the shell spacetimes is selected from all
spherically symmetric shell spacetimes as described in the previous
subsection, by fixing the value, $E$, say, of the Schwarzschild mass $E_-$ in
their internal subspacetimes. Then the spacetime ${\mathcal M}_1$ is a fixed,
complete Kruskal spacetime and each shell spacetime defines a trajectory
$\Sigma_1$ in it. This trajectory satisfies Eq.\ (\ref{radeq}) and
(\ref{timeq}) with $\epsilon = -1$. In Ref.\ \cite{B-H} a proof is given that
each such trajectory belongs to a unique shell spacetime: the condition that
the shell spacetime satisfies the full set of Einstein equations determines
the mass $E_+$ of the right subspacetime ${\mathcal M}_+$ and the trajectory
$R_+(t)$, $T_+(t)$ in it up to a constant shift of $T_+(t)$. In this sense,
the shell dynamics can be reduced to a dynamics of a particle-like object on a
fixed two-dimensional spacetime ${\mathcal M}_1$. In Ref.\ \cite{B-H}, the
corresponding variational principle was shown to be uniquely determined (up to
a coordinate-dependent factor in front of the super-Hamiltonian) if the
super-Hamiltonian was required to be at most quadratic in momenta, and the
value of the momentum conserved due to the time-shift symmetry to be the
negative of the Schwarzschild mass of the external subspacetime ${\mathcal
  M}_+$, that is the total energy of the shell.

Let us describe this variational principle.  We will leave out the indices
`$-$' and `1' in refering to ${\mathcal M}_-$ and ${\mathcal M}_1$ within the
Warsaw approach. The coordinates $x^a$, $a = 0,1$ on ${\mathcal M}$ play the
role of the canonical coordinates of the system; the corresponding canonical
momenta are denoted by $p_a$, $a = 0,1$.  The action reads
\be 
  S_1 = \int dt\,(p_a\dot{x}^a - {\mathcal NC}_1),
\label{act1}
\ee
where $t$ is an arbitrary parameter along dynamical trajectories,
${\mathcal N}$ is a Lagrange multiplier and the super-Hamiltonian
${\mathcal C}_1$ 
is given by
\be
  {\mathcal C}_1 := \frac{1}{2} F(g^{ab}p_ap_b + M^2) - Wg^{ab}p_a\xi_b 
  - \frac{1}{2}W^2. 
\label{superH1}
\ee
Here, $g^{ab}(x)$ is the contravariant metric on ${\mathcal M}$ (observe that
the supermetric $Fg^{ab}$ is flat), $\xi^a(x)$ is
the Schwarzschild time-shift Killing vector on ${\mathcal M}$, $F =
-g^{ab}\xi_a\xi_b$, the potential $W(x)$ is defined by
\be
  W(R) := E - \frac{M^2(R)}{2R},
\label{pot}
\ee
and $M(R)$ is the mass function of the shell matter.

Ref.\ \cite{B-H} contains a proof that the variation principle (\ref{act1})
implies the radial equation (\ref{radeq}) and the time equation (\ref{timeq})
with $\epsilon = -1$. The time equation (\ref{timeq}) with $\epsilon =
+1$ must be added by hand.

We observe that the extended phase space is four-dimensional and that there is
one constraint. Hence, the system has just one physical degree of freedom;
this can be chosen to be, for example, the radius of the shell.

As one easily verifies,
\[
  \{{\mathcal C}_1,\xi^ap_a\} = 0.
\]
Thus, $\xi^ap_a$ is conserved. The value of $\xi^ap_a$ is $-E_+$, where $E_+$
is the total energy of the shell, or alternatively, the Schwarzschild mass of
the external subspacetime of the shell spacetime. Thus, $p_a$ cannot be
homogeneous in the velocities $\dot{x}^a$; this is the reason for the presence
of the vector potential $W\xi_a$ in the super-Hamiltonian ${\mathcal C}_1$.

To do calculations, we shall choose the Kruskal coordinates $U$ and $V$ for
$x^a$; let us recapitulate some properties of these coordinates. The spacetime
metric has the form
\[
  ds^2 = -\frac{(4E)^2}{\kappa\text{e}^\kappa}dU\,dV,
\]
where $\kappa = \kappa(-UV)$, and the function $\kappa(x)$ is defined in the
interval $(-1,\infty)$ by its inverse, 
\[
  \kappa^{-1}(x) := (x-1)\text{e}^x;
\]
this implies the identity
\[
  \kappa'(x) = \frac{1}{\kappa(x)\text{e}^{\kappa(x)}}.
\]
 
The transformation to the Schwarzschild coordinates is given by
\be
  R = 2E\kappa(-UV),\quad T = 2E\ln\left|\frac{V}{U}\right|;
\label{RTUV}
\ee
it follows that
\be
  F(R) = -\frac{UV}{\kappa\text{e}^{\kappa}} =
  \frac{\kappa - 1}{\kappa}.
\label{UV}
\ee

The time-shift Killing vector $\xi$ has the form
\[
  \xi = \frac{1}{4E}(-U\partial_U + V\partial_V).
\]

The variational principle (\ref{act1}) and (\ref{superH1}) for the
Warsaw approach reads in the Kruskal coordinates as follows:
\be
  S_1 = \int dt(P_U\dot{U} + P_V\dot{V} - {\mathcal NC}_1),
\label{act4}
\ee
where
\be
  {\mathcal C}_1 = \frac{UVP_UP_V}{2(2E)^2} + \frac{\kappa - 1}{2\kappa}M^2 +
  \frac{W}{4E}(UP_U - VP_V) - \frac{W^2}{2}.
\label{superH4}
\ee

Our definitions (\ref{superH1}), (\ref{superH4}) and (\ref{pot}) of ${\mathcal
  C}_1$ and $W$ differ from those given in Ref.\ \cite{B-H} by the factor 
$F$ so that the super-Hamiltonian ${\mathcal C}_1$ and the potential $W$ are
smooth everywhere. Using Eq.\ (\ref{superH4}) one can show that the function
${\mathcal C}_1$ has a non-zero gradient everywhere on the phase space with
the exception of the points with $U=V=0$ (whose projections to the
configuration space is the crossing point of the two horizons $R = 2E$).
Moreover, the constraint equation ${\mathcal C}_1 = 0$ has no finite momenta
solution there in the generic case $W(2E) \neq 0$. This is roughly a
consequence of the requirement that $\xi^ap_a = E_+$ and the fact that all
components of the vector $\xi$ vanish at the crossing point.  Of course, in
the special case of flat spacetime ($E_- = 0$) this problem is avoided. We
shall also show in Sec.\ \ref{sec:cart1} that this singularity is due to the
choice of the coordinates $x^a$ and $p_a$.

An analysis that is based on the Kruskal coordinates cannot include the two
special cases, $E_-=0$ and $E_+=0$, in which one of the subspacetimes is flat
because the Kruskal coordinates become badly singular in the limit
$E_\pm\rightarrow 0$. Especially the case $E_-=0$ is, however, important. We
give a short description of it in the Appendix.

\subsection{The Potsdam approach}
In this subsection, another approach to the shell dynamics is described. It
starts so to speak from `the first principles': in Ref.\ \cite{H-K}, general
Lagrangian and Hamiltonian formalisms for massive shell are developed,
starting from the Einstein-Hilbert and a fluid actions. The Hamiltonian
formalism is reduced in Ref.\ \cite{H} by spherical symmetry, using a
transformation of canonical variables invented by Kucha\v{r} \cite{K}. The
basic properties of the resulting constrained system are as follows. 

The canonical coordinates involved are four coordinates of the shell in the
left and the right subspacetimes ${\mathcal M}_\pm$. For example,
one can take the Kruskal coordinates $U_\pm$ and $V_\pm$. Another variable is
the co-called `Kruskal momentum' $P^\epsilon_K$ that was introduced in Ref.\
\cite{H}: 
\[
  P^\epsilon_K := R_\epsilon\,\text{arctanh}\frac{dX_\epsilon}{dZ_\epsilon},
\]
where $X_\epsilon:=-U_\epsilon+V_\epsilon$ and
$Z_\epsilon:=U_\epsilon+V_\epsilon$, so that the argument of the arctanh can
be considered as the `Kruskal velocity', that is the velocity of the shell
with respect to the Kruskal frame $(n_\epsilon(1,1),n_\epsilon(-1,1))$
($n_\epsilon$ is a suitable normalization factor). First,
\be
  P^\epsilon_K = \frac{R_\epsilon}{2}\ln\frac{dV_\epsilon}{dU_\epsilon};
\label{Kmom}
\ee
observe that $dV_\epsilon/dU_\epsilon \geq 0$ holds along each non-spacelike
curve. There is a relation of the Kruskal momentum to the proper-time
velocity $(dU_\epsilon/ds,dV_\epsilon/ds)$; it follows from Eq.\ (\ref{Kmom})
and the normalization condition,
\be
  -\frac{(4E_\epsilon)^2}{\kappa_\epsilon\text{e}^{\kappa_\epsilon}}
  \frac{dU_\epsilon}{ds}\frac{dV_\epsilon}{ds} = -1,  
\label{normal}
\ee
where $\kappa_\epsilon := \kappa(-U_\epsilon V_\epsilon)$, namely,
\bea 
  \frac{dU_\epsilon}{ds} & = &
  \frac{\sqrt{\kappa_\epsilon\text{e}^{\kappa_\epsilon}}}{4E_\epsilon}\, 
  \text{e}^{-P^\epsilon_K/R_\epsilon},  
\label{dTds} \\
  \frac{dV_\epsilon}{ds} & = &
        \frac{\sqrt{\kappa_\epsilon\text{e}^{\kappa_\epsilon}}}{4E_\epsilon}\,
                \text{e}^{P^\epsilon_K/R_\epsilon}. 
\label{dRds} 
\eea
The variational principle of the Potsdam approach has been written down
in terms of Kruskal coordinates $U_\epsilon$, $V_\epsilon$, Kruskal momenta
$P^\epsilon_K$ and the energies $E_\epsilon$ in Ref.\ \cite{H}:
\be
 S_2 = \int dt\left([-2E\kappa\dot{P}_K +
 E^2(\kappa+1)\text{e}^{-\kappa}(V\dot{U}-U\dot{V})] + 2\tilde{\nu}[E\kappa] -
 \nu{\mathcal C}_2\right),
\label{act2}
\ee
where
\be
  {\mathcal C}_2 :=
  \left[E\sqrt{\kappa}\text{e}^{-\kappa/2}(-U\text{e}^{P_K/R} +
  V\text{e}^{-P_K/R})\right] + M(E_+\kappa_+ + E_-\kappa_-),
\label{superH2}
\ee
the symbol $[f]$ denotes
the jump of the quantity $f$ across the shell, $[f] := f_+ - f_-$ and
$\bar{f}$ denotes the average of the values of $f$ from left and right:
$\bar{f}:= (f_++f_-)/2$.

Let us stress that the origin of the symplectic structure defined by the
Liouville form in action (\ref{act2}) lies in the form of the Einstein-Hilbert
Lagrangian. As concerns the constraints, there are three of them in all: the
two primaries, $R_+ - R_- = 0$ and ${\mathcal C}_2 = 0$, and one secondary,
$\chi = 0$, which can be obtained by variation with respect to $P_{K\pm}$
after employing the constraint $E_+\kappa_+ = E_-\kappa_-$:
\[
  \chi := \frac{\partial C_2}{\partial P^+_K} + \frac{\partial C_2}{\partial
  P^-_K}.
\]
We obtain 
\[
  \chi = -\left[\frac{1}{2\sqrt{\kappa\text{e}^\kappa}}\left(U\text{e}^{P_K/R}
        + V\text{e}^{-P_K/R}\right)\right].
\]
The pair $(\chi, R_+ - R_-)$ form the second-class part of the constraints
set (see Ref.\ \cite{H}).

Hence, we have an eight-dimensional extended phase space and three
constraints. This implies that there are two physical degrees of freedom; they
can e.g.\ be chosen as $E_-$ and $R_-$. This should be compared with the
Warsaw approach of the previous subsection, where the Schwarzschild mass
$E_-$ was just a parameter having vanishing Poisson brackets with everything.

The Kruskal metric is invariant with respect to the one-dimensional
transformation group $g_\lambda$, $\lambda \in (-\infty,\infty)$:
\be
  U \mapsto U\text{e}^\lambda,\quad V\mapsto V\text{e}^{-\lambda}.
\label{UVlambda}
\ee
From this isometry, an infinite-dimensional abelian group of symmetry for the
action (\ref{act2}) can be constructed as follows. First, it can transform the
coordinates $U_-$, $V_-$ of the left and $U_+$, $V_+$ of the right
subspacetimes independently. Second, this double transformation can be
performed for each value of the pair $(E_+,E_-)$ independently. Such a
transformation has the form
\be
  U_\epsilon \mapsto U_\epsilon\text{e}^{\lambda_\epsilon(E_\epsilon)},\quad
  V_\epsilon\mapsto V_\epsilon\text{e}^{-\lambda_\epsilon(E_\epsilon)},\quad
  \epsilon = \pm 1,
\label{sigma}
\ee
where $\lambda_-(E_-)$ and $\lambda_+(E_+)$ are two arbitrary (smooth) {\em 
functions}. Let us denote the transformation by $G[\lambda_-(E_-),
\lambda_+(E_+)]$. Clearly, $G$ preserves the products $U_\epsilon V_\epsilon$
and the two masses $E_\epsilon$\footnote{The origin of this symmetry is easy
  to understand: continuous symmetry groups are generated by perennials
  (constants of motion) ; $E_-$ and $E_+$ are perennials, and any functions
  $\Lambda(E_-)$ and $\Lambda(E_+)$ are also perennials. Thus, one obtains a
  symmetry depending on two arbitrary functions of one variable. In fact, the
  symmetry group is even larger: it is generated by all perennials of the form
  $\Lambda(E_+,E_-)$!}. 

The corresponding transformation of the Kruskal momenta is determined by Eq.\
(\ref{Kmom}):
\be
  P^\epsilon_K\mapsto P^\epsilon_K - R\lambda_\epsilon(E_\epsilon). 
\label{PKlambda}
\ee
Then, the functions $U_\epsilon\text{e}^{P^\epsilon_K/R_\epsilon}$ and
$V_\epsilon\text{e}^{-P^\epsilon_K/R_\epsilon}$ are invariants, and so are both
constraints, ${\mathcal C}_2$ and $R_+-R_-$. A little more difficult is to see
that the Liouville form of the action (\ref{act2}) changes at most by a
closed form.  Let us write out the transformation of just the
$\epsilon$-part of it leaving out the index $\epsilon$ 
\beann
  \lefteqn{-RdP_K + E^2(1+\kappa)\text{e}^{-\kappa}(VdU-UdV)
    \rightarrow} \\
  && -RdP_K + E^2(1+\kappa)\text{e}^{-\kappa}(VdU-UdV) + Rd(R\lambda)
  + 2E^2(1+\kappa)\text{e}^{-\kappa}UVd\lambda.
\eeann
Using Eq.\ (\ref{UV}) and $R=2E\kappa$, we obtain for the last two terms:
\[
  Rd(R\lambda) + 2E^2(1+\kappa)\text{e}^{-\kappa}UVd\lambda = R\lambda dR +
  \frac{R^2}{2}d\lambda + 2E^2\frac{d\lambda}{dE}dE, 
\]
which is a closed form.

The Kruskal spacetime also admits an isometry group of four elements with
generators $T_1$ and $T_2$, the two transformations that are obtained by
extending the Schwarzschild time reversal $T\mapsto -T$, either from the
quadrant $U<0$, $V>0$ or from $U>0$, $V>0$ to the whole of the Kruskal
manifold: 
\[
  T_1(U,V)=(-V,-U),\quad T_2(U,V)=(V,U).
\]
These transformations can be used to define symmetries of the variational
principle (\ref{act2}) as follows. Consider first $T_1$; it changes the time
orientation of the Kruskal spacetime. Thus, it must act in both Kruskal
subspacetimes of each shell spacetime simultaneously, or else the resulting
shell spacetime has no time orientation. We have, therefore
\[
  T_1(U_\epsilon,V_\epsilon) = (-V_\epsilon,-U_\epsilon).
\]
The action of this transformation on the momentum $P^\epsilon_K$ is determined
by Eq.\ (\ref{Kmom}):
\[
  P^\epsilon_K \mapsto -P^\epsilon_K.
\]
Finally, $E_\epsilon$ are clearly invariants. This defines a transformation in
the extended phase space that will be also denoted by $T_1$. Observe that the
constraints ${\mathcal C}_2$ and $R_+-R_-$ are invariant, but that the
Liouvile form in the action (\ref{act2}) changes sign. Thus, $T_1$ is an
anti-symplectic map (this is common for time reversals).

Next, consider $T_2$; it changes the space orientation exchanging right and
left. Thus, we have to define:
\[
  T_2(U_\epsilon,V_\epsilon) = (V_{-\epsilon},U_{-\epsilon}),
\]
and, similarly,
\[
  P^-_K \leftrightarrow -P^+_K,\quad E_- \leftrightarrow
  E_+.
\]
One can see immediately that both the constraints and the Liouville form are
invariant with respect to this transfromation.

\section{Cartan forms}
\label{sec:cartan}
\subsection{Reparametrization-invariant reduction method}
The two variational principles (\ref{act4}) and (\ref{act2}) seem to be as
different as one can imagine. Although they depend on some common variables
(because $U=U_-$ and $V=V_-$), the momenta are different in each case, and
the nature of the super-Hamiltonians (\ref{superH4}) and (\ref{superH2}) is
very different: ${\mathcal C}_1 =0$ is a condition on the norm of the
two-velocity of the shell in the left subspacetime, whereas ${\mathcal C}_2
=0$ relates the time-time-components of the second fundamental forms of the
shell in the left and right subspacetime (indeed, ${\mathcal C}_2 =0$ is a
component of the Israel equation).

It turns out that the simplest way to compare the two systems is to solve all
constraints thus reducing the systems completely. It is, however, the
kind of reduction that does not deparametrize the system: no choice of gauge
is necessary. The resulting Lagrangian is the value $\Theta(\dot{X})$ of
the so-called {\em Cartan form} $\Theta$ at the tangent vector $\dot{X}$ of a
trajectory, and so it is homogeneous in velocities. The variation of the
corresponding action,
\[
  S = \int dt\,\Theta(\dot{X}),
\]
leads to dynamical equations of the form
\be
  d\Theta(\dot{X},\delta X) = 0,\quad \forall \delta X.
\label{12.1}
\ee 
$d\Theta$ coincides with the presymplectic form
$\Omega(\cdot,\cdot)$ on the constraint surface of the system, and the
equation of motion (\ref{12.1}) simply says that $\dot{X}$ must lie in the
degeneracy subspace of $\Omega$ (for more about presymplectic forms see, e.g.\ 
Refs.\ \cite{K-T} and \cite{S}, cf.\ also \cite{perenn}).

\subsection{Warsaw description}
\label{sec:cart1}
In this subsection, we will reduce the Warsaw system. Since the
Potsdam approach was written down only for massive shells as yet, we
restrict ourselves to massive shells, $M(R) \neq 0$, in the rest of the paper.
We observe then that the constraint ${\mathcal C}_1 = 0$ can be easily solved,
if we express the momenta $P_U$ and $P_V$ by means of $P_K$ as follows. The
variation of the action (\ref{act4}) with respect to the momenta gives their
relation to the velocities:
\beann 
  P_U & = & \frac{2(2E)^2}{UV}\left(\frac{\dot{U}}{{\mathcal N}} -
  \frac{WU}{4E}\right), \\
P_V & = & \frac{2(2E)^2}{UV}\left(\frac{\dot{V}}{{\mathcal N}} +
  \frac{WV}{4E}\right).  \eeann 
If we insert these expressions for the momenta
into the constraint (\ref{superH4}), we obtain that
\[
  \frac{\dot{U}}{\mathcal N}\cdot\frac{\dot{V}}{\mathcal N} =
  \frac{(\kappa-1)^2\text{e}^\kappa}{\kappa}\cdot\frac{M^2}{(4E)^2}.
\]
A comparison with Eq.\ (\ref{normal}) and a use of Eqs.\ (\ref{dTds}) and
(\ref{dRds}) leads to  
\bea
  P_U & = & 2E\left(-\frac{M}{\sqrt{\kappa\text{e}^\kappa}}\text{e}^{P_K/R} +
    \frac{W}{U}\right), 
\label{PUPK} \\
  P_V & = & 2E\left(-\frac{M}{\sqrt{\kappa\text{e}^\kappa}}\text{e}^{-P_K/R} -
  \frac{W}{V}\right). 
\label{PVPK}
\eea 
One easily verifies that these expressions for $P_U$ and $P_V$ satisfy
the constraint (\ref{superH4}) identically. In this way, we have arrived at
the constraint manifold $\Gamma_1$ with the coordinates $U$, $V$ and $P_K$. On
this manifold, there are two important quantities: the pull-back $-E_+$ of the
conserved quantity $\xi^ap_a = 1/(4E)(-UP_U + VP_V)$ and the Cartan form.
Substituting for the momenta $P_U$ and $P_V$, we obtain: 
\be 
  E_+ = E + \frac{M(R)}{2\sqrt{\kappa\text{e}^\kappa}}\left(-U\text{e}^{P_K/R}
    + V\text{e}^{-P_K/R}\right) - 
\frac{M^2(R)}{2R},
\label{BUV}
\ee
where $R$ is an abbreviation for $2E\kappa(-UV)$. The Cartan form is
obtained, if we substitute for the momenta into the Liouville form of the
action (\ref{act4}):
\be
  \Theta_1 = 2EWd\left(\ln\left|\frac{U}{V}\right|\right) -
  \frac{2EM(R)}{\sqrt{\kappa\text{e}^\kappa}}\left(\text{e}^{P_K/R}dU + 
  \text{e}^{-P_K/R}dV\right).
\label{thet1UV}
\ee

One would expect that these objects would be singular at the point $U=V=0$
where the horizons cross, and indeed, the Cartan form (\ref{thet1UV}) diverges
even at both horizons $U=0$ and $V=0$. However, this singularity can be
removed by subtracting the differential of the function
\[
  2EW(2E)\ln\left|\frac{U}{V}\right|;
\]
in this way, we obtain an everywhere regular Cartan form
\[
  \Theta_{1r} = 2EW_r(VdU-UdV) -
  \frac{2EM}{\sqrt{\kappa\text{e}^\kappa}}\left(\text{e}^{P_K/R}dU + 
  \text{e}^{-P_K/R}dV\right), 
\]
where
\[
  W_r(-UV) := \frac{W(R) - W(2E)}{UV}
\]
is a smooth function. For example, with $M=$ const,
\[
  W_r = \frac{M^2}{4E\kappa\text{e}^\kappa}.
\]

The regularity of the constraint system requires, however, a slightly stronger
condition that just the smoothness of the Cartan form: the presymplectic form
$d\Theta_1=d\Theta_{1r}$ must be smooth and its degeneracy subspace must be
one-dimensional everywhere in $\Gamma_{1}$. Let us calculate the presymplectic
form; a straightforward but tedious procedure yields: 
\beann 
  d\Theta_1 & = & \Bigglb(\frac{2(2E)^2W_R}{\kappa\text{e}^\kappa}
    + \frac{2E}{\kappa\text{e}^\kappa}
  \left(\frac{M}{\sqrt{\kappa\text{e}^\kappa}}\right)_\kappa
  \left(-U\text{e}^{P_K/R} + V\text{e}^{-P_K/R}\right) \\
& + &
  \frac{MP_K}{\kappa^3\text{e}^\kappa\sqrt{\kappa\text{e}^\kappa}}
  \left(U\text{e}^{P_K/R} 
    + 
    V\text{e}^{-P_K/R}\right)\Biggrb)\,dU\wedge dV \\
& - & \frac{M}{\kappa\sqrt{\kappa\text{e}^\kappa}} \left(\text{e}^{P_K/R}
  dP_K\wedge dU - 
  \text{e}^{-P_K/R}dP_K\wedge dV\right), 
\eeann 
where the indices $R$ and
$\kappa$ denote partial derivatives with respect to the corresponding
variables.  An inspection shows that this form is smooth and {\em non-zero}
everywhere; this is sufficient for a two form on a three-dimensional manifold
to define a smooth one-dimensional degeneracy distribution. Thus, the
constraint system $(\Gamma_{1},\Theta_{1})$ is completely regular even at
$U=V=0$. The singularity of the variational principle (\ref{act1}) is due just
to the choice of the variables $x^a$ and $p_a$: the simplicity of the
super-Hamiltonian (\ref{superH1}) is traded for the singularity at $U=V=0$.
In any case, the constraint manifold $\Gamma_1$ is defined by the coordinates
$U$, $V$, and $P_K$ in the ranges
\[
   -UV \in (-1,\infty),\quad E_+ \in (0,\infty).
\]

\subsection{Symmetry and adapted coordinates}
In this section, we shall study the static symmetry of the Kruskal spacetime
and the corresponding symmetry of the system $(\Gamma_1,\Theta_1)$. We shall
use the symmetry to find some coordinate systems which will simplify the
subsequent calculations. 

Eqs.\ (\ref{UVlambda}), and (\ref{PKlambda}) for $\lambda=$ const, define a
transformation on $\Gamma_1$; this transformation leaves the function $E_+$
and the form $\Theta_1$ invariant. Thus, it is a symmetry of the constraint
system. From the discrete transformations, only $T_1$ survives; $T_2$ would
map our system to an equivalent one, which would be obtained if we based
the description on the trajectory in the right subspacetime ${\mathcal M}_2$.
It is quite natural that this much smaller transformation group remained from
the group $G[\lambda_+(E_+),\lambda_-(E_-)]$ here, because the right
subspacetime is missing and $E=$ const, so $\lambda(E)=$ const. We call this
transformation $G(\lambda)$.

The orbits of the group $G(\lambda)$ in $\Gamma_1$ are one-dimensional, and the
quotient $\Gamma_1/G(\lambda)$ is a two-dimensional manifold. It is useful to
introduce coordinate systems that are adapted to this quotient structure: two
coordinates that are constant along the group orbits, and another one that
transforms simply by the group.

\subsubsection{Coordinates $u$, $v$ and $\tilde{P}$} 
Let us define:  
\be
  u := \frac{U\text{e}^{P_K/R}}{\sqrt{\kappa\text{e}^\kappa}},\quad v :=
  \frac{V\text{e}^{-P_K/R}}{\sqrt{\kappa\text{e}^\kappa}}. 
\label{defuv}
\ee
The factor $1/\sqrt{\kappa\text{e}^\kappa}$ does not improve the
transformation properties, but it leads to additional simplifications: indeed,
all exponentials, square roots and transcendental Kruskal functions will
disappear. Futher, 
\be
  \tilde{P} := \frac{EP_K}{R},
\label{tildeP}
\ee
hence
\[
  \tilde{P} \mapsto \tilde{P} - E\lambda.
\]
Eq.\ (\ref{UV}) implies that
\be
  uv = -1 + \frac{2E}{R},
\label{uvF}
\ee
or
\be
  R = \frac{2E}{1+uv},
\label{er}
\ee
and this, together with Eq.\ (\ref{RTUV}) yield
\be
  \kappa = \frac{1}{1+uv}.
\label{kapa}
\ee

The objects $E_+$ and $\Theta_1$ can be transformed into these
coordinates with the results: 
\be
  E_+ = E - \frac{M(R)}{2}(u - v) -
  \frac{M^2(R)}{2R}, 
\label{Buv}
\ee
and 
\bea
  \Theta_1 & = & -\frac{EM^2(R)}{R}d\left(\ln\left|\frac{u}{v}\right|\right) -
  2EM(R)d(u+v) \nn \\ 
  & - & \frac{M(R)}{2}(u+v)\left(1 + \frac{2E}{R}\right)dR -
  4E_+d\tilde{P}, 
\label{thetuv}
\eea 
where $R$ is defined by Eq.\ (\ref{er}). 
\subsubsection{Coordinates $R$, $E_+$ and $\tilde{P}$} 
The function $E_+$ is not
only invariant with respect to the group $G(\lambda)$, but also an integral of
motion. Hence, projections of the dynamical trajectories to the quotient
$\Gamma_1/G(\lambda)$ are just the curves $E_+=$ const. It will prove 
advantageous, therefore, to transform to the coordinates $R$
and $E_+$ on the quotient. Of course, $R$ and $E_+$ are not everywhere regular
coordinates: clearly, at the maximal radii of bound trajectories, gradients of
both functions $R$ and $E_+$ vanish in the direction of the trajectories; we
can, however, still work with these coordinates in the rest of the quotient,
and match the results at the singular points. The transformation from $u$ and
$v$ to $E_+$ and $R$ are given by Eqs.\ (\ref{Buv}) and (\ref{er}). The
inverse transformation is obtained by solving these equations for $u$ and $v$:
\bea
  u_\omega & = & \frac{-Y+\omega\sqrt{Y^2-4X}}{2},
\label{uRB} \\
v_\omega & = & \frac{Y+\omega\sqrt{Y^2-4X}}{2},
\label{vRB}
\eea
where $\omega$ is a sign, whose significance will be established shortly, 
\beann
  X & := & 1 - \frac{2E}{R}, \\
  Y & := & 2\left(\frac{E_+-E}{M(R)} + \frac{M(R)}{2R}\right),
\eeann
and
\be
  Y^2-4X = -4V(R),
\label{discYX}
\ee
where $V(R)$ is given by
Eq.\ (\ref{potent}). Thus, if $V(R)$ vanishes, $\dot{R}$ must also
vanish and this happens only at the turning point of a bound trajectory.
The meaning of the sign $\omega$ in front of the square root is simple. As
\beann
  (-u+v)_\omega & = & Y, \\
  (u+v)_\omega & = & \omega\sqrt{Y^2-4X},
\eeann
it distinguishes the upper and lower half of the $u$-$v$-plane; the turning
points are lying at $u+v=0$ and the coordinates $E_+$ and $R$ are
not regular there.

\subsection{Potsdam description}
In this subsection, we reduce our second system completely. In analogy with
Eq.\ (\ref{defuv}), we define 
\be
  u_\epsilon := \frac{U_\epsilon\text{e}^{P^\epsilon_K/R_\epsilon}}
  {\sqrt{\kappa_\epsilon\text{e}^{\kappa_\epsilon}}},\quad v_\epsilon :=
  \frac{V_\epsilon\text{e}^{-P^\epsilon_{K}/R_\epsilon}}
  {\sqrt{\kappa_\epsilon\text{e}^{\kappa_\epsilon}}},
\label{defuv+}
\ee
and obtain the relations
\be
  u_\epsilon v_\epsilon =  -1 + \frac{2E_\epsilon}{R_\epsilon},\quad 
  R_\epsilon =  \frac{2E_\epsilon}{1 + u_\epsilon v_\epsilon},\quad 
  \kappa_\epsilon = \frac{1}{1 + u_\epsilon v_\epsilon}.
\label{kapa+}
\ee
The three constraints ${\mathcal C}_2 = 0$, $R_+ = R_-$ and $\chi = 0$
simplify greatly in these coordinates:
\beann
  {\mathcal C}_2 & = & \frac{1}{2}\left[R(-u+v)\right] + M(\bar{R}), \\
  \chi & = & \frac{1}{2}[u+v], \\
  R & = & R_+ = R_-.
\eeann
They can be solved immediately, either for $u_+$, $v_+$ and $E_+$:
\bea
  u_+ & = & u_- + \frac{M(R)}{R},
\label{u+-} \\
  v_+ & = & v_- - \frac{M(R)}{R},
\label{v+-} \\
  E_+ & = & E_- - \frac{M(R)}{2}(u_- - v_-) - \frac{M^2(R)}{2R},
\label{E+-}
\eea
or for $u_-$, $v_-$ and $E_-$:
\be
  E_- = E_+ + \frac{M(R)}{2}(u_+ - v_+) - \frac{M^2(R)}{2R}
\label{E-+}
\ee
(solutions for $u_-$ and $v_-$ are given by Eqs.\ (\ref{u+-}) and
(\ref{v+-})). 

From the definitions Eqs.\ (\ref{defuv+}), we obtain 
\be
  \ln\left|\frac{U_\epsilon}{V_\epsilon}\right| = -\frac{2P^\epsilon_K}{R} +
  \ln\left|\frac{u_\epsilon}{v_\epsilon}\right|. 
\label{U/V+} 
\ee

To calculate the Cartan form $\Theta_2$, we first express the Liouville form
of the action (\ref{act2}) in the variables $u_-$, $v_-$, $P^-_K$, $E_-$,
$u_+$, $v_+$, $P^+_K$ and $E_+$ using an analogy of Eq.\ (\ref{UV})
\[
  u_\epsilon v_\epsilon = (1-\kappa_\epsilon)\text{e}^{\kappa_\epsilon};
\]
after substituting for $E_+$, $u_+$ and $v_+$ and employing several
times the equations $R=2E_-\kappa_-=2E_+\kappa_+$ we obtain: 
\be
  \Theta_2 \equiv
  -\left[\left(\frac{R^2}{4}-E^2\right)
    d\left(\ln\left|\frac{u}{v}\right|\right) + 4Ed\tilde{P}\right],
\label{thet2uv}
\ee
where the sign `$\equiv$' suggests that we have added some closed forms, and 
\be
  \tilde{P}_\epsilon := \frac{E_\epsilon P^\epsilon_K}{R}.
\label{tilP}
\ee
The form is well-defined and smooth everywhere on the constraint manifold
$\Gamma_2$ that is defined by the ranges of the coordinates $u_-$, $v_-$,
$E_-$, $\tilde{P}_-$ and $\tilde{P}_+$:
\[
  u_-v_- \in (-1,\infty),\quad u_+v_+ \in (-1,\infty),\quad E_- \in
  (0,\infty),
\]
\[
  E_+ \in (0,\infty),\quad \tilde{P}_- \in (-\infty,\infty),\quad \tilde{P}_+
  \in (-\infty,\infty). 
\] 
 
The reduced action has the form
\[
  S_{2C} = \int dt\,L_{2C},
\]
where the Lagrangian is
\be 
  L_{2C} = -\left[\left(\frac{R^2}{4}-E^2\right)\left(\frac{\dot{u}}{u} -
      \frac{\dot{v}}{v}\right) + 4E\dot{\tilde{P}}\right].
\label{lagr2r}
\ee

The symmetry group of the system $(\Gamma_2, \Theta_2)$ contains the whole
infinite-di\-men\-sion\-al group $G[\lambda_-(E_-),\lambda_+(E_+)]$. In the
coordinates $E_-$, $u_-$, $v_-$, $\tilde{P}_-$ and $\tilde{P}_+$, this
transformations take on the from:
\[
  E_-\mapsto E_-,\quad u_-\mapsto u_-,\quad v_-\mapsto v_-,\quad
  \tilde{P}_-\mapsto\tilde{P}_- - E_-\lambda(E_-),\quad
  \tilde{P}_+\mapsto\tilde{P}_+ - E_+\lambda(E_+). 
\]
$\Theta_2$ changes by $4E_+d(E_+\lambda_+(E_+)) - 4E_-d(E_-\lambda(E_-))$,
which is obviously a closed form. From the discrete group, $T_1$
remains simple; it acts as follows:
\[
  E_-\mapsto E_-,\quad u_-\mapsto -u_-,\quad v_-\mapsto -v_-,\quad
  \tilde{P}_-\mapsto -\tilde{P}_-,\quad \tilde{P}_+\mapsto -\tilde{P}_+.
\]

\section{Exclusion of a cyclic coordinate}
\label{sec:excl}
If one compares the systems $(\Gamma_1,\Theta_1)$ and $(\Gamma_2,\Theta_2)$
the first difference that catches one's eye is that $\Gamma_2$ is five- and
$\Gamma_1$ only three-dimensional. In this section, we shall get rid of two
dimensions of $\Gamma_2$. The fact that the variable $E$ appears in
$(\Gamma_1,\Theta_1)$ as a parameter implies that we have, as the first step,
to select the submanifold $\Gamma_{2E} \subset \Gamma_2$ of constant
coordinate $E_-$ in $\Gamma_2$. As $E_-$ is a constant of motion, the dynamical
trajectories that intersect $\Gamma_{2E}$ all remain in $\Gamma_{2E}$.
However, the dimension of $\Gamma_{2E}$ is still larger by 1 than that of
$\Gamma_2$; moreover, the pull-back $\Theta_{2E}$ of $\Theta_2$ to
$\Gamma_{2E}$ defines a presymplectic form $d\Theta_{2E}$ that has a
two-dimensional space of degeneracy: it contains the vector
$\partial/\partial\tilde{P}_-$ in addition to the direction of motion. Thus, in
the second step, we have to construct the manifold $\Gamma_E :=
\Gamma_{2E}/\tilde{P}_-$; it is the quotient of $\Gamma_{2E}$ by the
$\tilde{P}_-$-curves (i.e.\ curves with $u_-=$ const, $v_-=$ const and
$\tilde{P}_+=$ const). A form $\Theta'_{2E}$ that differs from $\Theta_{2E}$
by a form closed on $\Gamma_{2E}$ does not contain the variable $\tilde{P}_-$;
it can, therefore, be pushed forward by the quotient projection $\pi$ to
$\Gamma_E$ giving a form that we call $\Theta_E$. $d\Theta_E$ is a
presymplectic form on $\Gamma_E$ with a one-dimensional degeneracy subspace
everywhere, and the integral manifolds of this subspaces coincide with the
projection to $\Gamma_E$ of the original dynamical trajectories. Observe that
we have to take a quotient so that $\tilde{P}_-$ remains arbitrary; we are not
allowed to choose a surface transversal to the $\tilde{P}_-$-curves instead,
defined for example by the equation $\tilde{P}_-=$ const, for $\tilde{P}_-$ is
not a gauge coordinate or an integral of motion (it satisfies a non-trivial
dynamical equation that would be violated by $\tilde{P}_-=$ const).

In terms of coordinates and dynamical equations, the construction is very
simple. The pull-back $\Theta_{2E}$ of $\Theta_2$ to $\Gamma_{2E}$ coincides
with $\Theta_2$ in the coordinates $u_-$, $v_-$, $\tilde{P}_-$ and
$\tilde{P}_+$; only $E_-$ changes its status: it becomes a constant
parameter. Then the term $4E_-d\tilde{P}_-$ is the differential of the function
$4E_-\tilde{P}_-$ on $\Gamma_{2E}$ and can be omitted because it does not
contribute to $d\Theta_{2E}$; thus, we end up with the form
\be
  \Theta'_{2E} =
  -\left[\left(\frac{R^2}{4}-E^2\right)
  d\left(\ln\left|\frac{u}{v}\right|\right)\right] - 4E_+d\tilde{P}_+;
\label{thet2e}
\ee
this form is independent of $\tilde{P}_-$ and so can be pushed forward to
$\Gamma_E$; in the coordinates $u_-$, $v_-$ and
$\tilde{P}_+$, the push-forward is given by the same expression as
$\Theta_{2E}$. Thus, the new action reads
\[
  S_E = \int dt\,L_E,
\]
where the Lagrangian is
\be
  L_E = -\left[\left(\frac{R^2}{4}-E^2\right)\left(\frac{\dot{u}}{u} -
      \frac{\dot{v}}{v}\right)\right] - 4E_+\dot{\tilde{P}}_+
\label{20.1}
\ee
on the space $\Gamma_E$ with coordinates $u_-$, $v_-$ and $\tilde{P}_+$. Let
us denote this three-di\-men\-sion\-al system by $(\Gamma_E,\Theta_{E})$. The
symmetry of this system is still infinitely-dimensional: it is the
transformation group $G[\lambda_+(E_+),\lambda)$ (the $\epsilon=+1$-part of
the transformation acts only on $\tilde{P}_+$ and $\lambda_-(E_-)$ is a
constant, because $E_-$ is) and the time reversal $T_1$.

Let us compare the dynamical equations obtained by varying the actions
$S_{2C}$ and $S_E$. The following observation is helpful: the part of the
Lagrangian (\ref{lagr2r}) that contains the variables $u_-$, $v_-$ and
$\tilde{P}_+$ coincides with the Lagrangian (\ref{20.1}). It follows that the
variations of action $S_{2C}$ with respect to the variables $u_-$, $v_-$ and
$\tilde{P}_+$ differ from the corresponding variations of the action $S_E$
only by terms that are proportional to $\dot{E}_-$; but $\dot{E}_-=0$ is the
dynamical equation obtained by varying the action $S_{2C}$ with respect to
$\tilde{P}_-$. Hence, the three dynamical equations of the system
$(\Gamma_2,\Theta_2)$ due to variation of $u_-$, $v_-$ and $\tilde{P}_-$, if
one sets $\dot{E}_-=0$ in them, are identical with the corresponding three
equations of the system $(\Gamma_E,\Theta_E)$. The equation $\dot{E}_-=0$ is
free in the system $(\Gamma_E,\Theta_E)$, because $E_-$ is a constant
parameter there. Finally, there is another equation in the system
$(\Gamma_2,\Theta_2)$, namely that due to variation of $E_-$; it can be
written in the form
\be
  4\dot{\tilde{P}}_- = -\frac{\partial L_E}{\partial E_-};
\label{missingeq}
\ee 
it cannot be obtained in the system $(\Gamma_E,\Theta_E)$, because
$\tilde{P}_-$ is the cyclic coordinate that has been eliminated, and the new
variation principle does not contain any information about it. We observe that
$\tilde{P}_-$ is `completely smeared' even classicaly, as any point of
$\Gamma_E$ is a whole $\tilde{P}_-$-curve, $\tilde{P}_- \in (-\infty,
+\infty)$, and the remaining variables in the system $(\Gamma_E,\Theta_{E})$,
namely $u_-$, $v_-$ and $\tilde{P}_+$ have vanishing Poisson brackets with
$E_-$ in the original Potsdam system.

This is in a nice correspondence with the quantum mechanics construted for the
two classical systems (if the factor-ordering problem is solved suitably): the
Hilbert space ${\mathcal H}$ of the system $(\Gamma_2,\Theta_2)$ can be
written as the orthogonal sum of the eigenspaces ${\mathcal H}_{E_-}$ of the
operator $\hat{E}_-$; $E_-$ is a point of its spectrum $\sigma(\hat{E}_-)$. On
the Hilbert space ${\mathcal H}_{E_-}$, only those elements of the algebra of
observables can act that commute with $\hat{E}_-$, and ${\mathcal H}_{E_-}$ is
the Hilbert space of the system $(\Gamma_E,\Theta_E)$.

We can even reconstruct a trajectory of the system $(\Gamma_2,\Theta_2)$ from
one of $(\Gamma_E,\Theta_E)$ using the equation (\ref{missingeq}) which must
simply be added by hand, as follows. Let $u_-(t)$, $v_-(t)$ and
$\tilde{P}_+(t)$ be a trajectory of $(\Gamma_E,\Theta_E)$ parametrized by an
arbitrary parameter $t$. Let us substitute these functions for $u_-$, $v_-$
and $\tilde{P}_+$ in Eq.\ (\ref{missingeq}). We obtain an equation of the form
\[
  \frac{d\tilde{P}_-}{dt} = -\frac{1}{4}\frac{\partial L_E}{\partial
  E_-}(u_-(t),v_-(t),\tilde{P}_+(t));  
\]
the integration will yield the function $\tilde{P}_-(t)$ depending on one
arbitrary constant (this constant is determined by an initial value of
$\tilde{P}_-$). Such a procedure is, in fact, equivalent to reconstructing the
manifold $\Gamma_2$ from $\Gamma_E$ by
\[
  \Gamma_2 = \Gamma_E \times (-\infty,+\infty)_{E_-} \times
  (-\infty,+\infty)_{\tilde{P}_-} 
\]
and adding the term $4E_-d\tilde{P}_-$ to $\Theta_E$. In the quantum
mechanics, this corresponds to defining
\[
  {\mathcal H} := \sum_{\sigma(\hat{E}_-)}\,\otimes_\bot{\mathcal H}_{E_-}
\]
(in general, one has to be careful about the spectrum of the conserved
quantity; this is here just $(-\infty,+\infty)$) and 
\[
  \hat{\tilde{P}}_- := \frac{\text{i}}{4}\frac{\partial}{\partial E_-};
\]
the $E_-$-dependence of the Hamiltonian of the system $(\Gamma_E,\Theta_E)$
will then lead to some time evolution of $\hat{\tilde{P}}_-$.

\section{Search for the equivalence map}
\label{sec:trans}
It remains to find a transformation between the system
$(\Gamma_E,\Theta_E)$ and $(\Gamma_1,\Theta_1)$ showing that they are
equivalent. This problem will be studied in the present section.

The basic properties of such a map (which is, in fact a `morphism of
presymplectic manifolds'), let us denote it by $\Phi$, are the
following 
\begin{enumerate}
\item $\Phi : \Gamma_1 \mapsto \Gamma_E$ is a diffeomorphism.
\item The pull-back $\Phi_*\Theta_E$ of $\Theta_E$ to $\Gamma_1$ differs from
  $\Theta_1$ by a closed form. This guarantees that the presymplectic forms
  coincide\footnote{To define 
  equivalent dynamics, two presymplectic forms just had to be proportional to
  each other with an arbitrary function on the constraint surface as the factor
  of proportionality. However, presymplectic forms contain more information:
  they define the Poisson brackets of perennials (that is, functions that are
  constant along the trajectories, cf.\ \cite{perenn}); thus, the factor must
  be equal to 1.}.
\label{p:cond}
\item If $f_1 : \Gamma_1 \mapsto \mathbf{R}$ and $f_E : \Gamma_E \mapsto
  \mathbf{R}$ are quantities with the same physical or geometrical significance
  in both systems, then $\Phi_*f_E = f_1$.
\end{enumerate}

Suppose that such a map exists. Let us express it by means of the coordinates
$(u,v,\tilde{P})$ on $(\Gamma_1,\Theta_1)$ and $(u_-,v_-,\tilde{P}_+)$ on
$(\Gamma_E,\Theta_E)$ :
\bea
  u_- & = & u_-\circ\Phi (u,v,\tilde{P}),
\label{phi-u} \\
  v_- & = & v_-\circ\Phi (u,v,\tilde{P}),
\label{phi-v} \\
  \tilde{P}_+ & = & \tilde{P}_+\circ\Phi (u,v,\tilde{P}).
\label{phi-P}
\eea 
The functions (\ref{phi-u})--(\ref{phi-P}) can be viewed as a
coordinate transformation and $u$, $v$ and $\tilde{P}$ as new coordinates on
$\Gamma_E$. This is an interpretation of $\Phi$ that avoids the following
paradox: $\tilde{P}_-$ is to be totally undetermined on $\Gamma_E$ (see the
previous section) and at the same time $\tilde{P}_-$ has the same physical
meaning as $\tilde{P}$, which is to be a well-defined function on $\Gamma_E$ by
Eqs.\ (\ref{phi-u})--(\ref{phi-P}). By considering Eq.\ (\ref{phi-P}) as a
coordinate transformation, we regard $\tilde{P}$ as a coordinate on the right
subspacetime, because $\tilde{P}_+$ and $\tilde{P}$ are in one-to-one relation
[see also Eq.\ (\ref{25.1}) and (\ref{25.2})] and $\tilde{P}_+$ {\em is} such a
coordinate.  

Another important point is that there is an infinity of different maps that
satisfy above requirements, if {\em one} exists. This follows from symmetry:
given a map $\Phi$, we can sandwich it between the symmetries to obtain
another one. The non-uniqueness of $\Phi$, however, leads to an uncertainty in
the coordinate $\tilde{P}_+$ (and so to that of the Schwarzschild coordinate
$T_+$) of the shell. The symmetry $G$ is, from the point of view of the
spacetime geometry of each shell spacetime, nothing but a remainder of the
original general covariance: the Schwarzschild or the Kruskal coordinates are
not uniquely determined by the geometry of a fixed Kruskal spacetime. To map
the dynamics of the system $(\Gamma_1,\Theta_1)$ to $(\Gamma_E,\Theta_E)$, a
particular $\Phi$ must be fixed. Which one does not seem to matter as long as
(classical) physical properties are concerned: a change in $\Phi$ amounts just
to relabeling the classical dynamical trajectories.

In a quantum theory such a relabeling may lead to a problem, however. As an
example, consider the unitary extension of the quantum shell dynamics
described in Ref.\ \cite{HKK}. There, one works with coordinates analogous to
$u$, $v$ and $\tilde{P}$ and one uses the simplicity of the Hamiltonian in
these coordinates to show the existence (if the rest mass of the shell is
smaller than Planck mass) and uniqueness of a self-adjoint extension of the
Hamiltonian. The eigenfunctions of the extended Hamiltonian are linear
combinations of contracting and expanding waves. One can construct wave
packets from them that are, at each time $T$, spacially concentrated around a
well-defined wave-function maximum so that one can plot the radius $R_M$ of
the maximum as a function of the time $T$. The function $R_M(T)$ diverges for
$T\rightarrow -\infty$, then decreases till, at some $T_0$, a minimum
$R_M(T_0) > 0$ of $R_M(T)$ is reached, and then increases again to infinity as
$T\rightarrow \infty$. For each wave packet one can define a time delay
$\Delta T$ between the departure of the packet at $R=\infty$ and the arrival
at $R=\infty$. Of course, $\Delta T$ is not the limit
$\lim_{R=\infty}[T_2(R)-T_1(R)]$, where $T_{1,2}(R)$ are the two solutions of
the equation $R_M(T)=R$, because the difference $T_2(R)-T_1(R)$ diverges as
$R\rightarrow\infty$; one defines $\Delta T$ e.g.\ by comparing the packet
`trajectory' with some standard scattering trajectory.

To interpret this scattering, we need, however, the function $R_M(T_+)$ rather
than $R_M(T)$. In principle, we can calculate the first from the second by
using some fixed map $\Phi$. The problem is that two different maps, $\Phi_1$
and $\Phi_2$, say, will then lead to two different $R_M(T_+)$ that imply in
turn two different time delays $\Delta T$. This is clear because contracting
(expanding) part of $R_M(T)$ lies in the subset of the phase space that
contains contracting (expanding) scattering trajectories and the difference
between $\Phi_1$ and $\Phi_2$ along the expanding trajectories is independent
from their difference along contracting trajectories.

\subsection{The differential equation for $\Phi$ in the general case}
\label{sec:diff}
In this subsection, we shall return to the classical theory and reformulate
the above requirements on $\Phi$ in the form of a differential equation. We
assume that $E>0$ and $E_+>0$; the special case $E=0$ is studied in the
Appendix.  We have two manifolds, $\Gamma_1$ with coordinates $u$, $v$, and
$\tilde{P}$ that satisfy the conditions $R>0$ and $E_+> 0$, and $\Gamma_E$
with the coordinates $u_-$, $v_-$, $\tilde{P}_+$ satisfying the same condition
(which is independent of $\tilde{P}_-$ or $\tilde{P}_+$). According to the
point 2,
\be
  d\left(\Phi_*\Theta_E - \Theta_1\right) = 0.
\label{24.1}
\ee
The functions $u$ and $u_-$ as well as $v$ and $v_-$ represent the same
physical quantities, so according to the point 3 the transformation
(\ref{phi-u})--(\ref{phi-P}) has to 
preserve them:  
\bea
  u_- \circ \Phi(u,v,\tilde{P}) & = & u,
\label{25.1} \\
  v_- \circ \Phi(u,v,\tilde{P}) & = & v.
\label{25.2}
\eea
Thus, the only non-trivial part of $\Phi$ is the transformation (\ref{phi-P});
let us denote the corresponding function by $\phi$: 
\be
  \phi(u,v,\tilde{P}) := \tilde{P}_+ \circ \Phi(u,v,\tilde{P}).
\label{25.3}
\ee
Substituting Eqs.\ (\ref{25.1})--(\ref{25.3}) into Eq.\ (\ref{24.1}), we
obtain 
\be
  d\left(\Theta_E|_{E_-=E,u_-=u,v_-=v,\tilde{P}_+=\phi} - \Theta_1\right) = 0.
\label{25.4}
\ee
This is a partial differential equation of the first order for the function
$\phi$; its characteristics are easily found to coincide with the curves
$E_+=$ const. We can, 
therefore, reduce this equation to an ordinary differential equation, if we
transform it to the coordinates $R$ and $E_+$. That is our next task. 

Let us define the functions $A$, $B$, $A'$ and $B'$ of $R$ and $E_+$ by
\bea
  \Theta_E|_{E_-=E,u_-=u,v_-=v,\tilde{P}_+=\phi} & = & AdR + BdE_+ -
  4E_+d\phi, 
\label{AC} \\
  \Theta_1 & = & A'dR + B'dE_+ - 4E_+d\tilde{P}.
\label{ACprim}
\eea
Substituting Eqs.\ (\ref{AC}) and (\ref{ACprim}) into Eq.\ (\ref{25.4}), we
obtain the equations:
\beann
  A_{E_+} -A'_{E_+} - 4\phi_R & = & B_R - B'_R, \\
  \phi_{\tilde{P}} & = & 1
\eeann
(the indices denote partial derivatives). These two equations are equivalent
to the system 
\bea
  \phi & = & \tilde{P} + \Delta(R,E_+),
\label{defD} \\
  \frac{\partial \Delta}{\partial R} & = & \frac{1}{4}(A_{E_+} - B_R -
  A'_{E_+} + B'_R). 
\label{diffD}
\eea
Eq.\ (\ref{diffD}) is the desired ordinary differential equation (for
the function $\Delta$). 

Let us work out the explicit $R$-$E_+$-dependence of
the right-hand side. We obtain from Eq.\ (\ref{thetuv}) and (\ref{thet2e}):
\beann
  A & = & \left(\frac{R^2}{4}-E^2\right)
  \left(\ln\left|\frac{u}{v}\right|\right)_R - 
  \left(\frac{R^2}{4}-E_+^2\right)
  \left(\ln\left|\frac{u_+}{v_+}\right|\right)_R, \\  
  B & = & \left(\frac{R^2}{4}-E^2\right)
  \left(\ln\left|\frac{u}{v}\right|\right)_{E_+} - 
  \left(\frac{R^2}{4}-E_+^2\right)
  \left(\ln\left|\frac{u_+}{v_+}\right|\right)_{E_+}, \\ 
  A' & = & -\frac{EM^2}{R}\left(\ln\left|\frac{u}{v}\right|\right)_R -
  2EM(u+v)_R - \frac{M}{2}(u+v)\left(1 + \frac{2E}{R}\right), \\
  B' & = & -\frac{EM^2}{R}\left(\ln\left|\frac{u}{v}\right|\right)_{E_+} -
  2EM(u+v)_{E_+}.
\eeann 
We find easily
\beann
  A_{E_+} -B_R & = & 2E_+\left(\ln\left|\frac{u}{v}\right|\right)_R -
  \frac{R}{2}\left(\ln\left|\frac{u}{v}\right|\right)_{E_+} +
  \frac{R}{2}\left(\ln\left|\frac{u_+}{v_+}\right|\right)_{E_+}, \\
  A'_{E_+} -B'_R & = &
  E\left(\frac{M^2}{R}\right)_R\left(\ln\left|\frac{u}{v}\right|\right)_{E_+} +
  2EM_R(u+v)_{E_+} \\
  & - & \frac{M}{2}\left(1 + \frac{2E}{R}\right)(u+v)_{E_+}. 
\eeann
Next, we have to express everything in terms of $R$ and $E_+$,
using Eqs.\ (\ref{uRB}) and (\ref{vRB}). As the first step, we derive the
following helpful equations: 
\beann
  u & = & -\frac{E_+-E}{M}-\frac{M}{2R} + \omega\frac{\sqrt{\mathcal
        P}}{MR}, 
  \\
  v & = & \frac{E_+-E}{M}+\frac{M}{2R} + \omega\frac{\sqrt{\mathcal
        P}}{MR}, 
  \\
  u_+ & = & -\frac{E_+-E}{M}+\frac{M}{2R} + \omega\frac{\sqrt{\mathcal
      P}}{MR}, \\
  v_+ & = & \frac{E_+-E}{M}-\frac{M}{2R} + \omega\frac{\sqrt{\mathcal
      P}}{MR},
\eeann
where 
\[
  {\mathcal P}(R) := -M^2(R)R^2V(R).
\]
A straightforward calculation then gives
\beann
  A_{E_+} - B_R & = & -\frac{\omega}{\sqrt{\mathcal P}}(R+2E_+)
  \left(E_+-E+\frac{M^2}{2R}\right) + \\
  && \frac{\omega M_R}{M\sqrt{\mathcal
      P}}\left((4E_+^2-4E_+E)R+2M^2E_+\right), \\    
  A'_{E_+} - B'_R & = & -\frac{\omega}{2R\sqrt{\mathcal P}}\left(2(E_+-E)R^2
    + (4E_+E-4E^2+M^2)R -2EM^2\right) + \\
  && \omega\frac{2EM_R}{M\sqrt{\mathcal
      P}}\left(2(E_+-E)R-M^2\right). 
\eeann
Collecting all pieces, we arrive at the differential equation for $\Delta$
in the form
\bea
  \frac{d\Delta}{dR} & = & \frac{\omega}{4\sqrt{\mathcal P}}
  \left(-2(E_+-E)^2+M^2(E_++E)\frac{1}{R}\right) \nn \\ 
  & + & \omega\frac{M_R}{M\sqrt{\mathcal P}}\left((E_+-E)^2R+M^2(E_++E)\right).
\label{diffDRB}
\eea
Observe that the upper and lower $u$-$v$-plane give the same curves, only
their orientation is different.

\subsection{Solution of the equation for $\Delta$ in the case of dust}
In the general case, we cannot say much about the solution of Eq.\
(\ref{diffDRB}), because the function $M(R)$ in it is an arbitrary (positive)
function. In the case of dust, $M=$ const, however, the equation is
readily solvable. In this case, we obtain
\be
  \frac{d\Delta}{dR} = \frac{\omega}{4\sqrt{\mathcal P}}
  \left(-2(E_+-E)^2+M^2(E_++E)\frac{1}{R}\right).
\label{dD}
\ee
Eq.\ (\ref{dD}) can be solved by elementary integration; the solution,
however, will change its form if the parameters $E_+$, $E$ and $M$ vary. Thus,
we obtain only a local form of the function $\Delta$. In the present
subsection, a careful discussion of the different cases is given. The results
of this discussion will be used in the next subsection where we try to match
the different cases smoothly together.

Let us write the solution of Eq.\ (\ref{dD}) in each of the half $u$-$v$-plane,
$\omega=\pm 1$, as follows:
\[
  \Delta^\omega = \Delta^\omega_0 + \Delta^\omega_1 + \Delta^\omega_2,
\]
where $\Delta^\omega_0=\Delta^\omega_0(E_+)$ is an arbitrary function of $E_+$
(integration constant---this is the non-uniqueness in $\Phi$ due to
the symmetry), 
\[
  \Delta^\omega_1 := -\omega\frac{(E_+-E)^2}{2}\int\frac{dR}{\sqrt{\mathcal
  P}}, 
\]
and 
\[
  \Delta^\omega_2 := \omega\frac{M^2(E_++E)}{4}\int\frac{dR}{R\sqrt{\mathcal
  P}}. 
\]
For dust, the function $\mathcal P$ becomes a quadratic polynomial of $R$:
\[
  {\mathcal P}(R) := \left((E_+-E)^2 - M^2\right) R^2 + (E_++E)M^2 R +
  \frac{M^4}{4}, 
\]
The discriminant of the quadratic equation ${\mathcal P}(R) = 0$ equals to
$M^4(4EE_+ + M^2)$; it is always positive, so there are two roots:
\[
  R_1 = -\frac{M^2}{2}\frac{1}{(E_++E) + \sqrt{4E_+E+M^2}} < 0,
\]
and 
\be
  R_2 = -\frac{M^2}{2}\frac{(E_++E) + \sqrt{4E_+E+M^2}}{(E_+-E)^2 - M^2}.
\label{R2}
\ee
$R_2$ is positive for $(E_+-E)^2-M^2<0$, that is $E_+ \in
(\text{Max}(0,E-M),E+M)$. Then ${\mathcal P}(R)>0$ in the interval $R \in
(0,R_2)$ and the trajectory is bound, with maximal radius $R_2$. For
$(E_+-E)^2-M^2\geq 0$, $R_2<R_1<0$, ${\mathcal P}(R)>0$ in the interval $R \in
(0,\infty)$ and the trajectory is unbound. 

To calculate the function $\Delta^\omega_1$, we have to distinguish three
cases. \\ 
\underline{$(E_+-E)^2-M^2>0$}. In terms of the variables $E$, $E_+$ and $M$,
this means that either $E_+ > E + M$ or $E_+ < E - M$. The corresponding
trajectories are the contracting and expanding scattering states. Then
\be 
  \Delta^\omega_1(R,E_+) = -\omega\frac{(E_+-E)^2}{2\sqrt{\alpha}}\,
  \text{arccosh}\frac{2\alpha R+M^2(E_++E)}{M^2\sqrt{4E_+E+M^2}},
\label{delt11}
\ee
where $\alpha := (E_+-E)^2-M^2$; we use the increasing branch of arccosh.
$\Delta^\omega_1$ is regular for all $R \in (0,\infty)$. The values at the
end-points are:
\[
  \Delta^\omega_1(0,E_+) = -\omega\frac{(E_+-E)^2}{2\sqrt{\alpha}}\,
  \text{arccosh}\frac{E_++E}{\sqrt{4E_+E+M^2}},
\] 
and $\Delta^\omega_1$ diverges logarithmically with $R \rightarrow \infty$ to
$-\omega\infty$. \\
\underline{$(E_+-E)^2-M^2=0$}. Then $R_1 < 0$, $R_2 \rightarrow \infty$ and
the polynomial 
${\mathcal P}(R) > 0$ for all $R \in (0,\infty)$. The trajectories are the
parabolic scattering states separating the scattering states from the bound
states. The integral is: 
\be 
  \Delta^\omega_1 = -\omega\frac{(E_+-E)^2}{2(E_++E)}\sqrt{4(E_++E)R+M^2}.
\label{delt12}
\ee
It is again regular for all $R \in (0,\infty)$. The values at the end-points
are:
\[
  \Delta^\omega_1(0,E_+) = -\omega\frac{(E_+-E)^2M}{2(E_++E)}
\]
and it diverges as $R^{1/2}$ for $R \rightarrow \infty$. \\
\underline{$(E_+-E)^2-M^2<0$}. In terms of the variables $E$, $E_+$ and $M$,
this means that $E_+\in (\text{Max}(0,E - M),E + M)$. Then $R_1 < 0 < R_2$. The
corresponding trajectories are the bound states and the integral is
\be
  \Delta^\omega_1(R,E_+) = -\omega\frac{(E_+-E)^2}{2\sqrt{-\alpha}}\,
  \text{arccos}\frac{2\alpha R+M^2(E_++E)}{M^2\sqrt{4E_+E+M^2}}
\label{delt13}
\ee
(we use the decreasing branch of arccos).
$\Delta^\omega_1$ is regular for all $R \in (0,R_2)$. The values at the
endpoints are 
\[
  \Delta^\omega_1(0,E_+) = -\omega\frac{(E_+-E)^2}{2\sqrt{-\alpha}}\,
  \text{arccos}\frac{E_++E}{\sqrt{4E_+E + M^2}},
\]
and 
\be
 \Delta^\omega_1(R_2,E_+) = -\omega\frac{(E_+-E)^2}{2\sqrt{-\alpha}}\,\pi.
\label{r1val}
\ee

The function $\Delta^\omega_2$ can be calculated immediately with the result
\be
  \Delta^\omega_2(R,E_+) = -\omega\frac{E_++E}{2}\ln\frac{2(E_++E)R + M^2 +
  2\sqrt{\mathcal P}}{2R\sqrt{4E_+E+M^2}},
\label{delt2}
\ee
which is valid for all values of $E_+$, $E$, $M$, and $R$ for which ${\mathcal
  P} > 0$. $\Delta^\omega_2$ diverges at $R \rightarrow 0$ for all
trajectories. For scattering trajectories,
\[
 \Delta^\omega_2(\infty,E_+) = -\omega\frac{E_++E}{2}\ln\frac{(E_++E) +
  \sqrt{\alpha}}{\sqrt{4E_+E+M^2}}
\] 
is finite. For bound trajectories, 
\be
 \Delta^\omega_2(R_2,E_+) = -\omega\frac{E_++E}{2}\ln\frac{(E_++E) +
  \sqrt{4E_+E+M^2}}{2}
\label{r2val}
\ee
is also finite and for the limit $R_2 \rightarrow \infty$ ($E_+\rightarrow E\pm
M$), we obtain 
\[
 \lim_{R_2=\infty}\Delta^\omega_2(R_2,E_+) = -\omega\frac{2E\pm M}{2}\ln(2E\pm
  M). 
\]

\subsection{Matching and patching}
\label{sec:match}
The function $\Delta(R,E_+)$ must be well-defined and smooth for all values
$E_+>0$ and $R>0$ of its variables, and this at all values $E>0$ and $M>0$ of
its parameters in order that it determines a map $\Phi$ with the required
properties. The `pieces' of $\Delta$ obtained in the previous section must,
therefore, be smoothly matched together. In the present subsection we prove
that this is impossible and give some discussion of the negative result.

Within the half planes $u+v<0$ ($\omega<0$) and $u+v>0$ ($\omega>0$), the
function $\Delta_2^\omega(R,E_+)$ is smooth everywhere, but $\Delta^\omega_1$
as a function of $E_+$ seems to be divergent at $(E_+-E)^2-M^2=0$. A routine
inspection in the complex plane reveals that $\Delta^\omega_1$ is, in fact, a
smooth function there for each $\omega$. Hence, the sum
$\Delta_1^\omega(R,E_+) + \Delta_2^\omega(R,E_+)$ is smooth inside each half
plane. The whole function $\Delta^\omega$ can, therefore, be made smooth by an
arbitrary smooth choice of $\Delta_0^\omega$. 

The main problem is the matching at the boundary $u+v=0$ between the two half
planes. Let us first study some properties of the curves $E_+=$ const in the
$u$-$v$-plane. The Kruskal coordinates $U$ and $V$ are both future oriented;
hence the past singularity $R=0$ lies in the quadrant $U<0$, $V<0$, and the
future one lies in the quadrant $U>0$, $V>0$. The transformation (\ref{defuv})
[or (\ref{defuv+})] preserves signs, so the past singularity lies in the
lower, and the future one in the upper half of the $u$-$v$-plane. Thus, in the
lower half plane ($u+v<0$), $R$ is increasing along all trajectories and in
the upper half plane ($u+v>0$), $R$ is decreasing. Observe that only bound
trajectories can cross over from one half plane to the other; scattering
trajectories are always imprisoned inside one half plane: the expanding
($\dot{R}>0$) in the lower and the contracting ($\dot{R}<0$) in the upper half
plane.

Let us try to extend $\Delta$ continuously across the boundary $u+v=0$ using
the remaining freedom in $\Delta^\omega_0(E_+)$. We shall split
$\Delta^\omega_0(E_+)$ into two terms, $\Delta^\omega_{01}(E_+)
+\Delta^\omega_{02}(E_+)$. In the lower half plane, $\Delta^-_1$ is increasing
with $R$ if $E_+$ is kept constant. If $E_+ \in (\text{Max}(0,E-M),E+M)$,
$\Delta^-_1$ assumes the value (\ref{r1val}) for $\omega = -1$ at the boundary
$u+v=0$. In the upper half plane $\Delta^+_1$ is decreasing with $R$ and it
reaches the boundary with the value (\ref{r1val}) for $\omega = +1$. Thus, we
can construct a continuous function by choosing $\Delta^-_{01}(E_+)=0$ and
\[
  \Delta^+_{01}(E_+) = \frac{(E_+-E)^2}{\sqrt{-\alpha}}\,\pi.
\]
The function $\Delta^+_1(R,E_+) + \Delta^+_{01}(E_+)$ in the upper half plane,
together with $\Delta^-_1(R,E_+)$ in the lower, define in fact a {\em smooth}
function at all points the lower half plane, at the boundary $u+v=0$, and in
the subset $E_+ \in (\text{Max}(0,E-M),E+M)$ of the upper half plane; let us
denote this set by ${\mathcal D}_-$. ${\mathcal D}_-$ consists of all points
at the expanding scattering and at the bound trajectories. However, there is
no continuous extension of this function to the rest of the upper half plane,
because it diverges at the points of the upper half plane satisfying
$(E_+-E)^2-M^2=0$, that is, at the contracting parabolic trajectories.

Let us turn to the function $\Delta^-_2(R,E_+)$ starting again in the lower
half plane. It is increasing from the value $-\infty$ at $R=0$ to a finite
value at $R=\infty$ along all scattering trajectories, and to a finite value
(\ref{r2val}) with $\omega=-1$ along bound trajectories at the boundary
$u+v=0$. The function (\ref{r2val}) has {\em finite} limits as $E_+$
approaches the values $E\pm M$. Thus, there is no problem to extend this
$\Delta^-_2(R,E_+)$ to the whole upper half plane. One has just to choose
$\Delta^-_{02}(E_+)=0$ and
\[
  \Delta^+_{02}(E_+) = (E_++E)\ln\frac{E_++E+\sqrt{4E_+E+M^2}}{2}
\]
everywhere. The two functions $\Delta^-_2(R,E_+)$ in the lower and
$\Delta^+_2(R,E_+) + \Delta^+_{02}(E_+)$ in the upper half plane form together
a smooth function in the whole $u$-$v$-plane.

It is also clear that no continuous choice of $\Delta_{01}^- + \Delta_{02}^-$
can remove the singularity in the upper half plane. Hence, any allowed choice
of $\Delta$ that is continuous in the lower half plane will necessarily
diverge at all points of the contracting parabolic trajectories ($E_+ = E \pm
M$) in the upper half plane. This result already means that there is no
function $\Delta$ that would be continuous everywhere. 

An analogous construction starting from the upper half plane leads to
analogous results: $\Delta_1$ can be made smooth only in the open subset
${\mathcal D}_+$ of the $u$-$v$-plane that contains points of all contracting
scattering and all bound trajectories. It diverges at all expanding parabolic
trajectories.  $\Delta_2$ can again be chosen smooth everywhere. Of course,
the two constructions end up with two different solutions: in ${\mathcal D}_-
\cap {\mathcal D}_+$, they differ by a function of $E_+$ that diverges at all
points satisfying $E_+ = E\pm M$.

The two constructions in the foregoing paragraphs deliver, in fact, two
maximal continuous extensions of our local solutions for $\Delta$. This two
maximal continuous extensions are of course also smooth (if $\Delta_0^\omega$
are chosen so), but it is the continuity that is lost at the boundary. Similar
conclusions can be drawn for the case $E=0$ or $E_+=0$, but the proof must use
different vartiables, because $u$ and $v$ can serve as two regular coordinates
only for $E>0$. The proof for the special case $E=0$ is sketched in the
Appendix.

As discussed at the beginning of the present section, the map $\Phi$ (if it
existed) can be considered as equivalence map (morphism) between the two
systems $(\Gamma_1,\Theta_1)$ and $(\Gamma_E,\Theta_E)$ or, alternatively, as
a definition of new coordinates, $u$, $v$, and $\tilde{P}$, for the system
$(\Gamma_E,\Theta_E)$. The fact that $\Phi$ exists only locally and is
non-unique suggests another interpretation: it is a pasting map between
different patches of a larger presymplectic manifold. Let us describe an
example of such a construction and see, whether it can be of any use or not.

We shall denote the two maximal extensions of $\Phi$ defined above with
$\Delta_0^- = 0$ or $\Delta_0^+ = 0$ by $\Phi_-$ and $\Phi_+$, respectively;
their domains in $\Gamma_1$ are ${\mathcal D}_-$ and ${\mathcal D}_+$
\footnote{Observe that the maximal extensions can be chosen differently,
  overlapping again at the bound trajectories, but the new domain ${\mathcal
    D}'_-$ containing points of all scattering trajectories expanding to the
  right or contracting from the left and the new ${\mathcal D}'_+$ containing
  points of all scattering trajectories that expand to the left or that
  contract from the right.}. Consider two copies of $(\Gamma_E,\Theta_E)$
denoted by $(\Gamma'_E,\Theta'_E)$ and $(\Gamma''_E,\Theta''_E)$, and one copy
of $(\Gamma_1,\Theta_1)$; let $\Phi'_+ : \Gamma_1 \mapsto \Gamma'_E$ has the
domain ${\mathcal D}_+ \subset \Gamma_1$ and let $\Phi''_- : \Gamma_1 \mapsto
\Gamma''_E$ has the domain ${\mathcal D}_- \subset \Gamma_1$. Let us paste
$\Gamma_1$ and $\Gamma'_E$ together along ${\mathcal D}_+$ and
$\Phi'_+({\mathcal D}_+)$ by $\Phi'_+$ and similarly $\Gamma_1$ and
$\Gamma''_E$ along ${\mathcal D}_-$ and $\Phi''_-({\mathcal D}_-)$ by
$\Phi''_-$. The result is a well-defined (possibly non-Hausdorff)
presymplectic manifold, which will be denoted by ($\Gamma,\Theta)$.

Problems arise, however, if one tries to find a physical interpretation of
this construction and considers the role and position of observers. An idea,
which seems reasonable, is that each dynamical system like $\Gamma_E$ or
$\Gamma_1$ is to describe, in an idealized manner, what a family of
communicating observers can do with the shell. They can send it contracting in
different dynamical states, observe its motion and also observe different
states of expanding shells. It seems further reasonable that such observers
could be placed somewhere in the right asymptotic regions of ${\mathcal M}_+$
and ${\mathcal M}_-$ (or left, but not both right and left, because right
observers cannot communicate with the left ones). Indeed, the observers are in
${\mathcal M}_-$ before they throw in a shell contracting from right, and then
they are in ${\mathcal M}_+$; analogously for shells expanding to right, the
observers are first in ${\mathcal M}_+$ and, after the shell passes, in
${\mathcal M}_-$. In this sense, the right asymptotic regions of all spacetime
shell solutions of one system are to be considered as identical.

Next consider the pasting of three systems $\Gamma'_E$, $\Gamma_1$ and
$\Gamma''_E$. Nothing seems to hinder us in making the assumption that each of
the three dynamical systems has its own observer family of the above kind
before the pasting. It is, therefore, conceivable that some of the three
families remain distinct after the pasting. Then the observers of $\Gamma'_E$
might send in a shell and this shell disappears behind a horizon for these
observers, but appears during its motion somewhere else, where it can be
observed by another of the three families. Let us look to see what happens
with the families if we perform the pasting.

It is a pasting of {\em phase spaces}; but such a pasting implies also that
points of {\em spacetimes} will be identified. Indeed, let us chose any point
$(u'_-,v'_-,\tilde{P}'_+) \in \Gamma'_E$ that lies at a bound trajectory.
$(u'_-,v'_-,\tilde{P}'_+)$ determines via Eqs.\ (\ref{kapa+}) and (\ref{E+-})
the Schwarzschild mass of the right subspacetime ${\mathcal M}'_+$, via Eqs.\ 
(\ref{u+-}), (\ref{v+-}), (\ref{tilP}) and (\ref{defuv+}) a point
$(U'_+,V'_+)$ of ${\mathcal M}'_+$, where the shell is, as well as the
four-velocity of the shell at $(U'_+,V'_+)$ represented by $P^{\prime +}_K$
through Eq.\ (\ref{Kmom}). By $\Phi^{\prime -1}_+$, the point
$(u'_-,v'_-,\tilde{P}'_+)$ is identified with $(u,v,\tilde{P}) \in \Gamma_1$
and this, in turn by $\Phi''_-$ with the point $(u''_-,v''_-,\tilde{P}''_+)
\in \Gamma''_E$. Again, $(u''_-,v''_-,\tilde{P}''_+)$ determines the mass
$E''_+ = E'_+$ of ${\mathcal M}''_+$, the point $(U''_+,V''_+)$ of ${\mathcal
  M}''_+$ and a four-velocity $P^{\prime\prime +}_K$ at $(U''_+,V''_+)$.

It seems, therefore, that the points $(U'_+,V'_+)$ and $(U''_+,V''_+)$ must be
identical after the pasting, because they lie at the trajectory of one and the
same shell. In fact, one can easily see that all points of the trajectory then
lie in these particular spacetimes ${\mathcal M}'_+$ and ${\mathcal M}''_+$
and must be identical in pairs. Thus, ${\mathcal M}'_+$ and ${\mathcal M}''_+$
have a one-dimensional set of points in common. This is very strange {\em
  unless} the spacetimes ${\mathcal M}'_+$ and ${\mathcal M}''_+$ are
themselves identical.  In fact, there is only one isometry mapping ${\mathcal
  M}'_+$ onto ${\mathcal M}''_+$ and preserving the trajectory points. It is
natural to assume that such an identification is performed in all cases where
the above construction is viable; that is, ${\mathcal M}'_+$ and ${\mathcal
  M}''_+$ are identical for all $E'_+ = E''_+ \in (E_- - M,E_- +M)$. Notice
that there is no analogous argument for ${\mathcal M}'_-$ and ${\mathcal
  M}''_-$, because a point of $\Gamma'_E$ (or $\Gamma''_E$) does not determine
a unique point of ${\mathcal M}'_+$ (or ${\mathcal M}''_+$).

It seems to follow that the pasting leads to identification of all three
families of observers. Indeed, if the observers of the system $\Gamma'_E$
throw in a shell in a bound state (remember that there are bound states
passing through any radius, and, moreover, `lower' bound shells can in any
case be arraged indirectly by the asymptotic observers) the same must be done
at the same time and radius by each of the other families. This is
`disappointing' but not desastrous. However, if the observers of the system
$\Gamma'_E$ throw in a shell in a scattering state, then the same must be done
by the observers of $\Gamma_1$, but it {\em must not} be done by the observers
of $\Gamma''_E$, because contracting scattering states of $\Gamma'_E$ are
distinct from those of $\Gamma''_E$ according to the pasting procedure. This
seem to be a paradox; I have not found any way out of it as yet. Thus, the
pasting does not seem to work.

To summarize: The map $\Phi$ that realizes the equivalence between the systems
$(\Gamma_1,\Theta_1)$ and $(\Gamma_E,\Theta_E)$ is afflicted with two
problems. On one hand, a map satisfying all requirements 1--3 at the beginning
of Sec.\ \ref{sec:trans} does not exist, at least for the case of dust. One
might be able to weaken the continuity in some cautious way and still
preserve the physical content, but this is only a speculation that must be
studied. On the other hand, the map $\Phi$ is not unique; this does not seem
to lead to any serious difficulty as long as just the classical theory is
concerned, but it can be a handicap for the self-adjoint extension methods in
the quantum theory.

\subsection*{Acknowledgements}
Discussions with K.~Kucha\v{r} and G.~Lavrelashvili are thankfully
acknowledged.

\section*{Appendix} 
In this Appendix, we shall sketch the main steps of the line of reasoning in
terms of Schwarzschild coordinates so that the important special case $E_-=0$
of flat shell interior can be dealth with.

The Warsaw approach starts in this case from the action of the form
\be
  S_1 = \int dt\,(p_T\dot{T} + p_R\dot{R} - {\mathcal NC}_1),
\label{act3}
\ee
where $T$ and $R$ are coordinates on a two-dimensional Minkowski
half-spacetime $R>0$ and
\be
  {\mathcal C}_1 =  -\frac{1}{2}\left(-p_T +
      \frac{M^2(R)}{2R}\right)^2 +\frac{1}{2} p_R^2 
    + \frac{1}{2}M^2(R).
\label{superH3}
\ee
In order to reduce this action to the Cartan form, we introduce the
Schwarzschild (here: Minkowski) momentum $P_S$ by
\be
  P_{S} :=
  R\,\text{arctanh}\frac{dR}{dT}.
\label{PM}
\ee
The constraint ${\mathcal C}_1=0$ is then identically satisfied by
\beann
  p_T & = & -M(R)\,\text{cosh}\frac{P_S}{R} + \frac{M^2(R)}{2R}, \\ 
  p_R & = & M(R)\,\text{sinh}\frac{P_S}{R},
\eeann
and the Cartan form becomes
\be
  \Theta_1 = -M(R)\left(\text{cosh}\frac{P_S}{R}-\frac{M(R)}{2R}\right)dT + 
             M(R)\,\text{sinh}\frac{P_S}{R}dR.
\label{thetS}
\ee

The Potsdam approach in the Schwarzschild coordinates (Ref.\ \cite{H})
is awkward, because there are 16 disjoint ranges of validity of these
coordinates (4 quadrants for each Kruskal subspacetime). We shall need four
sign functions $a_\epsilon$ and $b_\epsilon$, $\epsilon = \pm 1$ and an
abbreviation $\text{sh}_ax$ in order that we can catch all 16 quadrants by a
single formula. The definitions are: $a_\epsilon := \text{sgn}F_\epsilon$,
where $F_\epsilon := 1-2E_\epsilon/R$, $b_\epsilon := +1$ in the past and
$b_\epsilon := -1$ in the future of the Kruskal {\em event} horizon in
${\mathcal M}_\epsilon$ and
\[
  \text{sh}_ax := \frac{\text{e}^x +a\text{e}^{-x}}{2}.
\]
The momentum $P^\epsilon_{S}$ conjugate to $R_\epsilon$ is determined by the `Schwarzschild velocity'
$dR/dT$ of the shell as follows (cf.\ \cite{H}): 
\be 
  P^\epsilon_{S} :=
  R_\epsilon\,\text{arctanh}\left(\frac{1}{F_\epsilon}
  \frac{dR_\epsilon}{dT_\epsilon}\right)^{a_\epsilon},
\label{PS}
\ee 
where $dR_\epsilon/dT_\epsilon$ is the derivative of the Schwarzschild
radius $R_\epsilon$ with respect to the Schwarzschild time $T_\epsilon$ along
the shell from the $\epsilon$-side.
The action reads
\be
  S_2 = \int dt\,\left(\left[P_{S}\dot{R} - E\dot{T}\right]
    + \tilde{\nu}(R_+ - R_-) - \nu {\mathcal C}_2\right),
\label{act5}
\ee
where $\tilde{\nu}$ and $\nu$ are Lagrange multipliers and 
\be
  {\mathcal C}_2 := \left[bR\sqrt{|F|}\,\text{sh}_{a}\frac{P_{S}}{R}\right] +
  M(\bar{R}). 
\label{superH5}
\ee
There is a secondary constraint, $\chi = 0$, where
\[
  \chi := \left(\frac{\partial {\mathcal
  C}_2}{\partial\bar{P_S}}\right)_{[P_S]}, 
\] 
explicitly, 
\[
  \chi = \left[b\sqrt{|F|}\,\text{sh}_{-a}\frac{P_{S}}{R}\right].
\]

The three constraints ${\mathcal C}_2 = 0$, $R_+ = R_-$ and $\chi = 0$ can be
rewritten as the following system of equations for $R_+$, $E_+$, $P_+$ $a_+$
and $b_+$:
\[
  R_+ = R,\quad a_+ = \text{sgn}F_+,
\]
\[
  b_+\sqrt{|F_+|}\,\text{sh}_{a_+}\frac{P^+_{S}}{R} =
  \text{cosh}\frac{P^-_S}{R} - \frac{M(R)}{R},
\]
\[
  b_+\sqrt{|F_+|}\,\text{sh}_{-a_+}\frac{P^+_{S}}{R} =
  \text{sinh}\frac{P^-_S}{R};
\]
we have left out the index ``$-$'' and we have already set $E=0$, $a_-=+1$,
$b_- = +1$ and $F_- = 1$ everywhere. It follows immediately that 
\be
  F_+ = \left(\text{cosh}\frac{P^-_S}{R} - \frac{M(R)}{R}\right)^2 -
  \text{sinh}^2\frac{P^-_S}{R},
\label{16.1}
\ee 
and 
\be
  \text{tanh}\frac{P^+_S}{R} = \left(\frac{\text{sinh}
      \frac{P^-_S}{R}}{\text{cosh}\frac{P^-_S}{R} -
      \frac{M(R)}{R}}\right)^{a_+}. 
\label{16.2}
\ee
Eq.\ (\ref{16.1}) can be rewritten in the form
\be
  F_+ = Q_1Q_2,
\label{16.3}
\ee
where
\bea
  Q_1 & := & 1 - \frac{M(R)}{R}\,\text{e}^{-P^-_S/R}, 
\label{16.4} \\
  Q_2 & := & 1 - \frac{M(R_S)}{R}\,\text{e}^{P^-_S/R}, 
\label{16.5}
\eea
are two important abbreviations. We also define the following sign functions:
\[
  q_1 := \text{sgn}Q_1,\quad q_2 := \text{sgn}Q_2.
\]
Then, Eq.\ (\ref{16.3}) implies
\be
  a_+ = q_1q_2,
\label{16.6}
\ee
and
\[
  E_+ = \frac{R}{2}(1 - Q_1Q_2).
\]
Explicitly,
\be
  E_+ = E_- + M(R)\text{cosh}\frac{P^-_S}{R} - \frac{M^2(R)}{2R}.
\label{17.1}
\ee

Let us turn to Eq.\ (\ref{16.2}). We use the following simple identity
\[
  \text{arctanh}x^a = \frac{1}{2}\ln\left(a\frac{1+x}{1-x}\right)
\]
that holds for $a = \pm 1$ and all $|x| < 1$. It implies that
\[
  P^+_{S} = \frac{R}{2}\ln\left(a_+\frac{Q_1}{Q_2}\right) + P^-_S.
\]
However,
\[
  a_+\frac{Q_1}{Q_2} = \left|\frac{Q_1}{Q_2}\right|
\]
because of Eq.\ (\ref{16.6}) so that
\[
  P^+_{S} = \frac{R}{2}\ln\left|\frac{Q_1}{Q_2}\right| + P^-_S
\]
or equivalently
\be
  \text{e}^{P^+_{S}/R} =
  \sqrt{\left|\frac{Q_1}{Q_2}\right|}\text{e}^{P^-_S/R}. 
\label{17.2}
\ee

As for $b_+$, we just substitute Eq.\ (\ref{17.2}) for e$^{\pm P^+_S/R}$
and Eq.\ (\ref{16.3}) for $F_+$ into the equation $\chi = 0$; this leads
to
\[
  b_+\left(|Q_1|\text{e}^{P^-_S/R} - a_+|Q_2|\text{e}^{-P^-_S/R}\right)
  = 2\text{sinh}\frac{P^-_S}{R}
\]
Using the definitions of $q_1$, $q_2$, $Q_1$ and $Q_2$, we obtain easily
\be
  b_+ = q_1.
\label{17.3}
\ee
From Eqs.\ (\ref{16.6}) and (\ref{17.3}), it follows that $Q_1$ changes sign
at the ``black hole'' and $Q_2$ at the ``white
hole'' horizon of ${\mathcal M}_+$.

Subsituting Eq.\ (\ref{17.1}) for $E_+$, Eq.\ (\ref{17.2}) for $P^+_{S}$ and
$R$ for $R_+$ and $R_-$ into the variational principle (\ref{act3}), we obtain
the reduced action with the Cartan form $\Theta_E$ given by
\be
  \Theta_2 := \frac{R}{2}\ln\left|\frac{Q_1}{Q_2}\right|dR -
  M(R)\left(\text{cosh}\frac{P^-_S}{R} - \frac{M(R)}{2R}\right)dT_+.
\label{17.4}
\ee 
The cyclic coordinate $T_-$ is automatically excluded, because it is
contained only in one term of the form $E_-dT_-$ and $E_-=0$. The manifold
$\Gamma_E$ is determined by the inequalities
\[
  R > 0,\quad R \neq 2E_+,\quad E_+ \geq 0,
\]
that must be satisfied by the coordinates $R$, $P^-_S$, and $T_+$; it
consists of four open disjoint submanifolds.

Finally, we have to show that there is a transformation
\[
  T_+ = \tilde{\phi}(T,R,P_S),
\]
such that $\Theta_E$ becomes to $\Theta_1$ plus possibly a closed form,
if we substitute $\tilde{\phi}$ for $T_+$. In an analogous way as in Sec.\
\ref{sec:diff}, we obtain
\[
  \tilde{\phi}(T,R,P_S) = T + \tilde{\Delta}(R,P_S),
\]
and
\be
  \frac{d\tilde{\Delta}}{dR} =
  \text{sgn}P_S\,\frac{2E_+R^2-M^2(R)R}{2(R-2E_+)\sqrt{{\mathcal P}(R)}} -
  \text{sgn}P_S\,\frac{2E_+R+M^2(R)}{2\sqrt{{\mathcal P}(R)}}.
\label{difftilD}
\ee 
In the case of dust, $M(R)=$ const, the differential equation (\ref{difftilD})
reduces to 
\be
 \frac{d\tilde{\Delta}}{dR} = \text{sgn}P_S\,\frac{2E_+^2 -
   M^2}{\sqrt{{\mathcal P}(R)}} 
  + \text{sgn}P_S\,\frac{E_+(4E_+^2 - M^2)}{(R-2E_+)\sqrt{{\mathcal P}(R)}}, 
\label{dtilD}
\ee
where
\[
  {\mathcal P}(R) := (E_+^2-M^2)R^2 + M^2E_+R + M^4/4.
\]

The integration of Eq.\ (\ref{dtilD}) is completely analogous to that of
(\ref{dD}) and the results are also analogous, only the divergence of
$\Delta_2$ at $R=0$ is shifted to $R=2E_+$. This divergence of
$\tilde{\Delta}_2$ at $R=2E_+$ does not lead to any problem; it originates
from the singularity of the coordinate $T_+$ on the Kruskal spacetime
${\mathcal M}_+$ rather that from any geometrical effect.

\end{document}